\documentclass[nofigure]{pasj00}
\Received{2012/3/17}
\Accepted{2012/7/20}
\SetRunningHead{Moeen Moghaddas et al.}{Shear tensor and dynamics of relativistic disks}

\usepackage{times}

\begin{document}

\title{Shear tensor and dynamics of relativistic accretion disks around rotating black holes}
\author{Mahboobe Moeen Moghaddas} %
\affil{Department of Physics, School of Sciences, Ferdowsi University of Mashhad, Mashhad, 91775-1436, Iran}
\email{mahboobemoen@gmail.com}

\author{Jamshid Ghanbari}
\affil{Department of Physics, School of Sciences, Ferdowsi University of Mashhad, Mashhad, 91775-1436, Iran\\ Department of Physics, Khayam Institute of Higher Education, Mashhad, Iran}\email{ghanbari@ferdowsi.um.ac.ir}
\and
\author{Ahmad Ghodsi}
\affil{Department of Physics, School of Sciences, Ferdowsi University of Mashhad, Mashhad, 91775-1436, Iran}\email{ahmad@ipm.ir}
KeyWords{: black hole accretion disks, relativistic disks, accretion disks, shear tensor, hydrodynamic.}
\maketitle

\begin{abstract}
In this paper we solve the hydrodynamical equations of optically
thin, steady state accretion disks around Kerr black holes. Here,
fully general relativistic equations are used. We use a new method
to calculate the shear tensor in the LNRF (Locally Non-Rotating
Frame), BLF (Boyer-Lindquist Frame) and FRF (Fluid Rest Frame). We
show that two components of shear tensor in the FRF are nonzero (in
previous works only one nonzero component was assumed). We can use
these tensors in usual transonic solutions  and usual causal
viscosity, but we derive solutions analytically by some
simplifications. Then we can
calculate the four velocity and density in all frames such as the
LNRF, BLF and FRF.
\end{abstract}

\section{Introduction}

Accretion disks are important in several astrophysical systems. They
can be found around Young Stellar Objects(YSO),
around compact stellar objects in our galaxy and around several super-massive
black holes in Active Galactic Nuclei(AGN).

In super massive accretion disks the mass of black hole is
$(10^5-10^9)M_{\bigodot}$. To study such disks we use the general
relativity with relativistic hydrodynamics in Kerr metric background
geometry. In relativistic Navier-Stokes fluid we have stress-energy
tensor which is related to viscosity and is the cause of
redistribution of energy and momentum in fluid. This tensor is
defined by the four velocity and metric. In the previous studies,
the relativistic and stationary solutions of standard black hole
disks were solved. Lasota (1994) was the first who wrote down
slim-disk equations which include relativistic effects. He also
assumed that only r-$\phi$ component of the stress tensor is
nonzero. He used a special form for this component, which was
followed by Abramowicz et al(1996 and 1997). Chakrabarti (1996)
derived the transonic solutions of thick and thin disks for a weak
viscosity. He assumed similar form for viscosity as Lasota (1994).
Then Manmoto (2000) derived the global two temperatures structure of
advection-dominated accretion flows (ADAFs) numerically by using
full relativistic hydrodynamical equations including the energy
equations for the ions and electrons.

Papaloizou \& Szuskiewicz (1994) introduced a phenomenological and
non-relativistic equation for the evolution of viscous stress tensor
(causal viscosity) which has been used by many authors. Gammie \&
Popham (1998) and Popham \& Gammie (1998) solved  ADAFs with
relativistic causal viscosity. They used the Boyer-Lindquist
coordinates. \newpage
Takahashi (2007b) solved the equations of relativistic
disks in the Kerr-Schild coordinates by using the relativistic
causal viscosity. In the papers of Gammie \&
Popham and Takahashi they assumed that in the FRF, only the $r-\phi$
component of shear viscosity is nonzero, then they used the
transformation tensors to derive the components of shear stress tensor in the
Boyer-Lindquist or Kerr-Schild frames.

In present study, we concentrate on the stationary axisymmetric
accretion flow in the equatorial plane. We use a new method to
calculate the shear tensor and azimuthal velocity of fluids in the
locally non-rotating frame (LNRF) by using Keplerian angular
velocity. We derive two kinds of shear tensors in LNRF and BLF; in
the first one the direction of fluid rotation is the same as that of
black hole ($\Omega^{+}$) and in the second one the direction of
fluid rotation is in opposite to that of black hole($\Omega^{-}$).
We calculate the components of the shear tensor for two kinds of
fluids in LNRF, BLF and FRF, these calculations show that in FRF
there are two nonzero components ($r-t$ and $r-\phi$ components).
But in previous papers, the only nonzero component in FRF was
$r-\phi$ component. The $r-t$ component results from the
relativistic calculations of the shear tensor which changes some
components of the four velocity. Then, by using these shear tensor
components we calculate the four velocity in LNRF, BLF and density
in all frames .

This paper's agenda is as follows. We introduce the metric and
reference frame in $\S$2. In $\S$3 basic equations are given. In
$\S$4 the shear tensor is calculated in FRF, LNRF and BLF. We derive
four velocity in LNRF and BLF in $\S$5 and the influence of two
important parameters on density and four velocity can be seen in
this section. Also, the influence of the $r-t$ component of shear
tensor can be seen in this section. Summery and conclusion are given
in $\S$6.
\section{Metric, Reference Frame }
\subsection{Back Ground Metric}
For back ground geometry, we use the Boyer-Lindquist coordinates of
the rotating black hole space time. In the Boyer-Lindquist
coordinates, the Kerr metric is:
\begin{eqnarray}\label{1}
ds^{2}=&&g_{\alpha
\beta}dx^{\alpha}dx^{\beta}=-\alpha^{2}dt^{2}+\gamma_{ij}(dx^{i}+\beta^{i}dt)(dx^{j}\nonumber\\&&+\beta^{j}dt),
\end{eqnarray}
where $i,j=r,\theta$ and $\phi$. Nonzero components of the lapse
function $\alpha$, the shift vector $\beta^{i}$ and the spatial matrix
$\gamma_{ij}$ are given in the geometric units as:
\begin{eqnarray}\label{2}
&\alpha&=\sqrt{\frac{\Sigma\Delta}{A}},\qquad \beta^{\phi}=-\omega,\qquad \gamma_{rr}=\frac{\Sigma}{\Delta},\nonumber\\
&\gamma_{\theta\theta}&=\Sigma,\qquad \gamma_{\phi\phi}=\frac{Asin^{2}\theta}{\Sigma}.
\end{eqnarray}
Here, we use geometric mass $m=GM/c^{2}$,
$\Sigma=r^{2}+a^{2}\cos^{2}\theta$, $\Delta=r^{2}-2Mr+a^{2}$ and
$A=\Sigma\Delta+2mr(r^{2}+a^{2})$. The position of outer and inner
horizons, $r\pm$, are calculated by inserting $\Delta=0$ to get
$r\pm=m\pm(m^2-a^2)^{1/2}$. The angular velocity of the frame
dragging due to the black hole rotation is
$\omega=-g_{t\phi}/g_{\phi\phi}=\frac{2mar}{A}$, where M is the
black hole mass, G is the gravitational constant and c is the speed
of light and the angular momentum of the black hole, J, is described
as:
\begin{equation}\label{1'}
a=Jc/GM^{2},
\end{equation}
where $-1<a<1$.

Similar to Gammie \& Popham (1998), we set $G=M=c=1$ for basic
scalings. The nonzero components of metric, $g_{\mu\nu}$, and its
inverse, $g^{\mu\nu}$ are calculated in Appendix 1.
\subsection{Reference Frame}
In our study, we use three reference frames. The first one is the
Boyer-Lindquist frame (BLF) based on the Boyer-Lindquist coordinates
describing the metric, in which, our calculations are done. The
second one is the locally non-rotating reference frame (LNRF) which
is formed by observers with a future-directed unit vector orthogonal
to $t=constant$. By using the Boyer-Lindquist coordinates, the LNRF
observer is moving with the angular velocity of frame dragging
($\omega$). The third frame is fluid rest frame (FRF), an
orthonormal tetrad basis carried by observers moving along the
fluid.

The physical quantities measured in LNRF are described by using the
hat such as $u^{\hat{\mu}}$ and $u_{\hat{\mu}}$ and in FRF by using
parentheses such as $u^{(\mu)}$ and $u_{(\mu)}$. The transformation
matrixes of FRF, LNRF and BLF are given in Appendix 2 (Bardeen 1970;
Bardeen, Press \& Teukolsky 1972; Frolov \& Novikov 1998).

\section{Basic Equations}
The basic equations for the relativistic hydrodynamics are the
baryon-mass conservation $(\rho u^{\mu})_{;\mu}=0$ and the energy
momentum conservation; $T^{\mu\nu}_{;\nu}=0$, where $\rho$ is the
rest-mass density and $T^{\mu\nu}$ is the energy-momentum tensor.
Basic dynamical equations except the baryon mass conservation are
calculated from the energy-momentum tensor, $T^{\mu\nu}$. We use the
energy-momentum tensor written as:
\begin{equation}\label{4}
T^{\mu\nu}=\rho\eta
u^{\mu}u^{\nu}+pg^{\mu\nu}+t^{\mu\nu}+q^{\mu}u^{\nu}+q^{\nu}u^{\mu},
\end{equation}
where p is the pressure, $\eta=(\rho+u+p)/\rho$ is
the relativistic enthalpy, u is the internal energy, $t^{\mu\nu}$
is the viscous stress-energy tensor and $q^{\mu}$ is the heat-flux
four vector. The relativistic Navier-Stokes shear stress, $t^{\mu\nu}$, is written as (Misner, Thorne \& Wheeler 1973):
\begin{equation}\label{5}
t^{\mu\nu}=-2\lambda\sigma^{\mu\nu}-\zeta\Theta h^{\mu\nu},
\end{equation}
where $\lambda$ is the coefficient of dynamical viscosity, $\zeta$
is the coefficient of bulk viscosity, $h^{\mu\nu}=g^{\mu\nu}+u^{\mu}u^{\nu}$ is the projection
tensor, $\Theta=u^{\gamma}_{;\gamma}$  is the expansion of the
fluid world line, and $\sigma^{\mu\nu}$ is the shear tensor of the
fluid which is calculated as:
\begin{equation}\label{6}
\sigma_{\mu\nu}=\frac{1}{2}(u_{\mu;\nu}+u_{\nu;\mu}+a_{\mu}u_{\nu}+a_{\nu}u_{\mu})-\frac{1}{3}\Theta
h_{\mu\nu},
\end{equation}
where $a_{\mu}=u_{\mu;\gamma}u^{\gamma}$ is the four acceleration.

We study a stationary, axisymmetric and equatorially symmetric
global accretion flow  in the equatorial plane, i.e., we assume
$u_{\theta}=0$. We also assume that the effects of the bulk
viscosity and heat-flux four vector is negligible. In the following
sections, we derive the basic equations 
using the vertical averaging procedures
around the equatorial plane which were derived in, e.g., Gammie \&
Popham (1998).
\subsection{Mass conservation}
The equation for the baryon mass conservation is written as:
\begin{equation}\label{7}
(\rho
u^{\mu})_{;\mu}=\frac{1}{\sqrt{-g}}(\sqrt{-g}\rho u^{\mu})_{,\mu}=0,
\end{equation}
where $u^{\mu}$ is the four velocity and $\sqrt{-g}=r^2$. By averaging the physical quantities
around the equatorial plane and assuming constant $\dot{M}$ (the
mass-accretion rate ), we can write equation (\ref{7}) as follows:
\begin{equation}\label{8}
(4\pi H_{\theta}r^2 \rho u^{r})_{,r}=0  \Rightarrow  -4\pi
H_{\theta}r^2 \rho u^{r}=\dot{M},
\end{equation}
where $H_{\theta}$ is half-thickness of the accretion disk in
$\theta$ direction. If we normalize the rest-mass density , $\rho$,
by setting $\dot{M}=1$, for calculating global structure of accretion
flow, we have:
\begin{equation}\label{11}
-4\pi H_{\theta}\rho u^{r}r^{2}=1,
\end{equation}
also, by differentiating equation (\ref{11}) we have:
\begin{equation}\label{12}
\frac{d\ln H_{\theta}}{dr}+\frac{2}{r}+\frac{d\ln
u^{r}}{dr}+\frac{d\ln \rho}{dr}=0.
\end{equation}
\subsection{Killing Vector}
Two specific killing vectors of Kerr metric are $\varepsilon
^{\mu}_{t}=(1,0,0,0)$ and $\varepsilon^{\mu}_{\phi}=(0,0,0,1)$,
which are used to derive the disk equations. At first, angular
momentum conservation can be derived by $\varepsilon^{\mu}_{\phi}$
\begin{equation}\label{13}
(T^{\nu}_{\mu}\varepsilon^{\mu}_{\phi})_{;\nu}=0\Rightarrow
(T^{\nu}_{\phi})_{;\nu}=0.
\end{equation}
By vertically averaging equation (\ref{13}) we have:
\begin{equation}\label{14}
\frac{1}{r^2}(r^2 T^{r}_{\phi})_{,r}=0\Rightarrow \dot{M}\eta l-4\pi
H_{\theta}r^2 t^{r}_{\phi}=\dot{M}j,
\end{equation}
where $\dot{M}j$ is the total inward flux of angular momentum. This
equation is similar to the angular momentum equation of Gammie \&
Popham (1998).

Another equation which can be obtained by using $\varepsilon^{\mu}_{t}$ is:
\begin{equation}\label{15}
(T^{\nu}_{\mu}\varepsilon^{\mu}_{t})_{;\nu}=0\Rightarrow
(T^{\nu}_{t})_{;\nu}=0.
\end{equation}
Similarly, by vertically averaging and inserting constant $\dot{E}$, we have:
\begin{equation}\label{16}
4\pi H_{\theta}r^2((p+\rho+u)u_{t} u^{r}+t^{r}_{t})=\dot{E}.
\end{equation}
This equation expresses the constancy of mass-energy flux $\dot{E}$
in terms of the radius; $\dot{E}$ is the actual rate of change of the
black hole mass. If the fluid is cold and slow at large radiuses, then
$\dot{E}\approx \dot{M}$ (Gammie \& Popham 1998).
\subsection{Energy Equation}
The equation of local energy conservation is obtained from $u_{\mu}T^{\mu\nu}_{;\nu}=0 $, as follows:
\begin{eqnarray}\label{17}
&&-u^{r}\frac{d(\rho+u)}{dr}-(\rho+u+p)\Theta+(\rho+u+p)u^{\nu}
u^{\mu}_{;\nu} u_{\mu}\nonumber\\&&+t^{\mu\nu}_{;\nu}u_{\mu}=0,
 \end{eqnarray}
where $\Theta$ is defined as:
\begin{equation}\label{19}
\Theta=u^{\nu}_{;\nu}.
\end{equation}
And from equation (\ref{7}) we have:
\begin{equation}\label{18}
(\rho u^{\nu})_{;\nu}=0\Rightarrow\rho _{;\nu}u^{\nu}+\rho
u^{\nu}_{;\nu}=\frac{d\rho}{dr}u^r+\rho u^{\nu}_{;\nu}=0.
\end{equation}
Therefore the energy equation can be written as:
\begin{equation}\label{20}
u^{r}(\frac{du}{dr}-\frac{u+p}{\rho}\frac{d\rho}{dr})=(\rho+u+p)a^{\mu}
 u_{\mu}+t^{\mu\nu}_{;\nu}u_{\mu}.
\end{equation}
In this paper we do not derive the temperature, pressure and
internal energy so that, we will not use this equation. If we want
to derive these variables we must use a state equation and use the
relation of shear tensor components.
\section{Shear Tensor}
Lasota (1994) used the $r-\phi$ component of shear stress tensor as follows:
\begin{equation}\label{21}
t^{r}_{\phi}=-\nu\rho\frac{A^{3/2}\Delta^{1/2}\gamma_{\phi}^{3}}{r^{5}}\frac{d\Omega}{dr},
\end{equation}
\begin{equation}\label{22}
\gamma_{\phi}\equiv (1-(v^{\hat{\phi}})^{2})^{-1/2},
\end{equation}
where $v^{\hat{\phi}}$ is the azimuthal component of velocity in the
LNRF which is introduced by Manmoto (2000):
\begin{equation}\label{23}
     v^{\hat{\phi}}=\frac{A}{r^{2}\Delta^{1/2}}\tilde{\Omega},
\end{equation}
where $\tilde{\Omega}$ is defined as:
\begin{equation}\label{24}
    \tilde{\Omega}=\Omega-\omega.
\end{equation}
Abramowicz et al (1996 and 1997), Manmoto (2000) and others also
used equation (\ref{21}) for stress tensor. In 1994, Papaloizou \&
Szuskiewicz introduced a non-relativistic causal viscosity which was
used in relativistic form by  Peitz \& Apple (1997a,b), Gammie \&
Popham (1998)and Popham \& Gammie (1998) in Kerr metric and
Takahashi (2007b) in Kerr-Schild metric. They assumed that in the
FRF all components vanish, except $t_{r\phi}=t_{\phi r}$. The
$t_{r\phi}=S$ can be introduced in relativistic form as:
\begin{equation}\label{25}
 \frac{DS}{D\tau}=-\frac{S-S_{0}}{\tau_{r}},
\end{equation}
where $D/D\tau=u^{\mu}( )_{;\mu}$ and $S_{0}$ is the equilibrium
value of the stress tensor and $\tau_{r}$ is the relaxation time
scale. Therefore, in steady state we have :
\begin{equation}\label{26}
u^{r}\frac{dS}{dr}=-\frac{S-S_{0}}{\tau_{r}}.
\end{equation}
We use a new method to derive the shear tensor approximately. We
calculate $v^{\hat{\phi}}$ from equation (\ref{23}) in LNRF
assuming $u^{r}=0$ (the azimuthal velocity is much greater than
accretion velocity or radial velocity, except very close to the
inner edge, and for fluids with a small viscosity). In order to simplify,
we assume $\Omega=\Omega_{k}$, where Keplerian angular
velocity $\Omega_{k}$, is defined as:
\begin{equation}\label{10}
\Omega_{k}^{\pm}=\pm\frac{M}{r^{3/2}\pm aM^{1/2}},
\end{equation}
 therefore, for azimuthal velocity we have:
\begin{equation}\label{27}
v^{\hat{\phi}}=\frac{A}{r^{2}\Delta^{1/2}}(\frac{\pm1}{r^{3/2}\pm a}-\frac{2ar}{A}).
\end{equation}
First we calculate the
$u^{\hat{t}}=\hat{\gamma}=\sqrt{\frac{1}{1-(v^{\hat{\phi}})^2}}$.


We can calculate four velocity in LNRF (Appendix 3). The
four velocity for $\Omega^{-}$ (direction of fluid
rotation is opposite to the black hole rotation) is:
\begin{eqnarray}
&&u_{\hat{\mu}}=\nonumber\\&& (\frac{-r\sqrt{\Delta}(r^\frac{3}{2}-a)}{\sqrt{r^4(r^\frac{3}{2}-a)^2\Delta-A^2-4a^2r^2(r^\frac{3}{2}-a)^2-4arA(r^\frac{3}{2}-a)}}\nonumber\\&&,0,0,\nonumber\\&&\frac{-(r^{3}+ra^{2}+2ar^{\frac{3}{2}})}{\sqrt{r^4(r^\frac{3}{2}-a)^2\Delta-A^2-4a^2r^2(r^\frac{3}{2}-a)^-4arA(r^\frac{3}{2}-a)}}).\nonumber\\
\end{eqnarray}
Calculations show that in LNRF $\Gamma^{\hat{\alpha}}_{\hat{\mu}\hat{\nu}}=0$ and $\Theta=0$. The
shear tensor can be calculated by inserting the four
velocity of LNRF in equation (\ref{6}). The nonzero components of $u^{-}_{\hat{\mu};\hat{\nu}}$, $a^{-}_{\hat{\mu}}$($a_{\hat{\mu}}=u_{\hat{\mu};\hat{\gamma}}u^{\hat{\gamma}}$) and $\sigma_{\hat{\mu}\hat{\nu}}$ for $\Omega^{-}$ in LNRF are:
\begin{eqnarray}\label{'}
&&u^{-}_{\hat{t};\hat{r}}=\frac{1}{2\sqrt{\Delta}B^{-}}(+24r^6a^3+12ar^8-36r^5a^3-18ar^7\nonumber\\&&+12a^5r^4+6a^5r^3+r^\frac{19}{2}+3a^6r^\frac{7}{2}+2a^6r^\frac{5}{2}-36r^\frac{11}{2}a^2\nonumber\\&&+5r^\frac{15}{2}a^2+10r^\frac{13}{2}a^2+7r^\frac{11}{2}a^4+4r^\frac{9}{2}a^4+4r^\frac{7}{2}a^4),\nonumber\\
&&u^{-}_{\hat{\phi};\hat{r}}=\frac{1}{2\sqrt{\Delta}B^{-}}(+2r^\frac{11}{2}a^3+9ar^\frac{15}{2}+12r^\frac{9}{2}a^3-18ar^\frac{13}{2}\nonumber\\&&-3a^5r^\frac{7}{2}-2a^5r^\frac{5}{2}+r^9+18r^5a^2+4r^7a^2-20r^6a^2\nonumber\\&&+3r^5a^4-4r^4a^4-2r^3a^4),
\end{eqnarray}
\begin{eqnarray}
a^{-}_{\hat{t}}&=&u^{-}_{\hat{t};\hat{r}}u^{\hat{r}}=0,\qquad a^{-}_{\hat{r}}=u^{-}_{\hat{r};\hat{t}}u^{\hat{t}}+u^{-}_{\hat{r};\hat{\phi}}u^{\hat{\phi}}=0,\qquad\qquad\nonumber\\ a^{-}_{\hat{\phi}}&=&u^{-}_{\hat{\phi};\hat{r}}u^{\hat{r}}=0,
\end{eqnarray}
\begin{eqnarray}\label{02}
&&\sigma^{-}_{\hat{t}\hat{r}}=\sigma^{-}_{\hat{r}\hat{t}}=\frac{1}{4\sqrt{\Delta}B^{-}}(+24r^6a^3+12ar^8-36r^5a^3-18ar^7\nonumber\\&&+12a^5r^4+6a^5r^3+r^\frac{19}{2}+3a^6r^\frac{7}{2}+2a^6r^\frac{5}{2}-36r^\frac{11}{2}a^2\nonumber\\&&+5r^\frac{15}{2}a^2+10r^\frac{13}{2}a^2+7r^\frac{11}{2}a^4+4r^\frac{9}{2}a^4+4r^\frac{7}{2}a^4),\nonumber\\
&&\sigma^{-}_{\hat{r}\hat{\phi}}=\sigma^{-}_{\hat{\phi}\hat{r}}=\frac{1}{4\sqrt{\Delta}B^{-}}(+2r^\frac{11}{2}a^3+9ar^\frac{15}{2}+12r^\frac{9}{2}a^3+r^9\nonumber\\&&-18ar^\frac{13}{2}-3a^5r^\frac{7}{2}-2a^5r^\frac{5}{2}+18r^5a^2+4r^7a^2-20r^6a^2\nonumber\\&&+3r^5a^4-4r^4a^4-2r^3a^4),
\end{eqnarray}
where $B^{-}=(r^7-2r^\frac{11}{2}a-r^4a^2-3r^6-6r^3a^2+r^5a^2-2r^\frac{7}{2}a^3-4a^3r^{\frac{5}{2}})^\frac{3}{2}$.
Also for $\Omega^{+}$, following Appendix 3, the four
velocity is:
\begin{eqnarray}
&&u_{\hat{\mu}}=\nonumber\\&& (\frac{-r\sqrt{\Delta}(r^\frac{3}{2}+a)}{\sqrt{r^4(r^\frac{3}{2}+a)^2\Delta-A^2-4a^2r^2(r^\frac{3}{2}+a)^2+4arA(r^\frac{3}{2}+a)}}\nonumber\\&&,0,0\nonumber\\&&,\frac{r^{3}+ra^{2}-2ar^{\frac{3}{2}}}{\sqrt{r^4(r^\frac{3}{2}+a)^2\Delta-A^2-4a^2r^2(r^\frac{3}{2}+a)^2+4arA(r^\frac{3}{2}+a)}}).\nonumber\\
\end{eqnarray}
Therefore, the nonzero components of $u_{\hat{\mu};\hat{\nu}}$ and
$\sigma_{\hat{\mu}\hat{\nu}}$ for $\Omega^{+}$ are(similar to
$\Omega^{-}$, $a_{\hat{\mu}}=0$):
\begin{eqnarray}\label{''}
&&u^{+}_{\hat{t};\hat{r}}=\frac{1}{2\sqrt{\Delta}B^{+}}(-24r^6a^3-12ar^8+36r^5a^3+18ar^7\nonumber\\&&-12a^5r^4-6a^5r^3+r^\frac{19}{2}+3a^6r^\frac{7}{2}+2a^6r^\frac{5}{2}-36r^\frac{11}{2}a^2\nonumber\\&&+5r^\frac{15}{2}a^2+10r^\frac{13}{2}a^2+7r^\frac{11}{2}a^4+4r^\frac{9}{2}a^4+4r^\frac{7}{2}a^4),\nonumber\\
&&u^{+}_{\hat{\phi};\hat{r}}=\frac{-1}{2\sqrt{\Delta}B^{+}}(-2r^\frac{11}{2}a^3-9ar^\frac{15}{2}-12r^\frac{9}{2}a^3+18ar^\frac{13}{2}\nonumber\\&&+3a^5r^\frac{7}{2}+2a^5r^\frac{5}{2}+r^9+18r^5a^2+4r^7a^2-20r^6a^2\nonumber\\&&+3r^5a^4-4r^4a^4-2r^3a^4),
\end{eqnarray}
\begin{eqnarray}\label{01}
&&\sigma^{+}_{\hat{t}\hat{r}}=\sigma^{+}_{\hat{r}\hat{t}}=\frac{1}{4\sqrt{\Delta}B^{+}}(-24r^6a^3-12ar^8+36r^5a^3+18ar^7\nonumber\\&&-12a^5r^4-6a^5r^3+r^\frac{19}{2}+3a^6r^\frac{7}{2}+2a^6r^\frac{5}{2}-36r^\frac{11}{2}a^2\nonumber\\&&+5r^\frac{15}{2}a^2+10r^\frac{13}{2}a^2+7r^\frac{11}{2}a^4+4r^\frac{9}{2}a^4+4r^\frac{7}{2}a^4),\nonumber\\
&&\sigma^{+}_{\hat{r}\hat{\phi}}=\sigma^{+}_{\hat{\phi}\hat{r}}=\frac{-1}{4\sqrt{\Delta}B^{+}}(-2r^\frac{11}{2}a^3-9ar^\frac{15}{2}-12r^\frac{9}{2}a^3+r^9\nonumber\\&&+18ar^\frac{13}{2}+3a^5r^\frac{7}{2}+2a^5r^\frac{5}{2}+18r^5a^2+4r^7a^2-20r^6a^2\nonumber\\&&+3r^5a^4-4r^4a^4-2r^3a^4),
\end{eqnarray}

where $B^{+}=(r^7+2 r^\frac{11}{2}a-r^4 a^2-3 r^6-6 r^3 a^2+r^5 a^2+2 r^\frac{7}{2}a^3+4 a^3 r^{\frac{5}{2}})^{\frac{3}{2}}$.
The shear tensor in BLF can be derived from
$\sigma_{\alpha\beta}=e^{\hat{\mu}}_{\alpha}e^{\hat{\nu}}_{\beta}\sigma_{\hat{\mu}\hat{\nu}}$
in which $e^{\hat{\mu}}_{\alpha}$ and $e^{\hat{\nu}}_{\beta}$ are
the transformation tensors(Appendix 2). There are two nonzero
components of shear tensor in BLF, $\sigma_{tr}$ and
$\sigma_{r\phi}$, which are given in Appendix 4.
Also using transformation matrices, two nonzero components of shear
tensor in FRF can be calculated including $\sigma_{(t)(r)}$ and
$\sigma_{(r)(\phi)}$ which are given in Appendix 4 as well.

\section{Deriving four velocity }
Assuming special values for $\lambda$ (for example in Takahashi
(2007b), $\lambda=1.5,1.7,2.1,2.2,2.3,2.4$ and $2.5$), we can solve
equations (\ref{12}), (\ref{14}), (\ref{16}) and (\ref{20}) to
derive four velocity, density, pressure and etc, numerically. In
numerical solutions we must insert values of physical variables on
boundary conditions specially on horizon. Some of these boundary
conditions are nonphysical, therefore we want to derive analytical
solutions without any boundary conditions. We use some
non-relativistic relations to calculate four velocity and density in
LNRF and BLF analytically. At first, the $t_{r\phi}$ relation of
stress tensor in Takahashi (2007a) is used in a relativistic
disk,this relation is:
\begin{equation}\label{32}
t_{r\phi}=-\nu\rho r^2\frac{d\Omega}{dr}.
\end{equation}
Similar to Abramowicz et al.(1996) we assume $\eta=1$, therefore
from equation (\ref{5}) by zero bulk viscosity and
$\lambda=\rho\eta\nu$, we have:
\begin{equation}\label{33}
\sigma^{\pm}_{r\phi}=\frac{r^2}{2}\frac{d\Omega}{dr},
\end{equation}
therefore, $\Omega^{\pm}$ is
\begin{equation}\label{34}
\Omega^{\pm}=\int\frac{2}{r^2}\sigma^{\pm}_{r\phi}dr.
\end{equation}
For calculation the $\Omega^{\pm}$ we use equations \ref{02} and
\ref{01} with a simplification in $\frac{1}{B^{+}}$ and
$\frac{1}{B^{-}}$ as:
\begin{eqnarray}
&&\frac{1}{B^{+}}=(r^7+2 r^\frac{11}{2}a-r^4 a^2-3 r^6-6 r^3 a^2+r^5 a^2\nonumber\\&&+2 r^\frac{7}{2}a^3+4 a^3 r^\frac{5}{2})^{-\frac{3}{2}}\nonumber\\&&\approx\frac{1}{\sqrt{r^{21}}}(1+\frac{3}{2r^7}(3r^6-2ar^\frac{11}{2}-r^5a^2+r^4a^2+6r^3a^2\nonumber\\&&-r^5a^2-2r^\frac{7}{2}a^3-4a^3 r^\frac{5}{2})+\frac{15}{8r^{14}}(3 r^6+2 r^\frac{11}{2}a+r^4 a^2\nonumber\\&&+6 r^3 a^2-r^5 a^2+2 r^\frac{7}{2}a^3+4 a^3 r^{\frac{5}{2}})^2),
\end{eqnarray}
\begin{eqnarray}
&&\frac{1}{B^{-}}=(r^7-2 r^\frac{11}{2}a-r^4a^2-3r^6-6r^3a^2+r^5a^2\nonumber\\&&-2r^\frac{7}{2}a^3-4a^3 r^\frac{5}{2})^{-\frac{3}{2}}\nonumber\\&&\approx\frac{1}{\sqrt{r^{21}}}(1+\frac{3}{2r^7}(3r^6+2ar^\frac{11}{2}-r^5a^2+r^4a^2+6r^3a^2\nonumber\\&&-r^5a^2+2r^\frac{7}{2}a^3+4a^3 r^\frac{5}{2})
+\frac{15}{8r^{14}}(3 r^6-2 r^\frac{11}{2}a+r^4 a^2\nonumber\\&&+6 r^3 a^2-r^5 a^2-2 r^\frac{7}{2}a^3-4 a^3 r^{\frac{5}{2}})^2).
\end{eqnarray}
Influences of this approximation in shear tensor components are seen
in figure 1 for $\Omega^{+}$ and $a=.9$. Solid curves have no
simplifying and dotted curves are with this simplifying.
\begin{center}
\input{epsf}
\epsfxsize=3.2in \epsfysize=2.3in
\begin{figure*}
\centerline{\epsffile{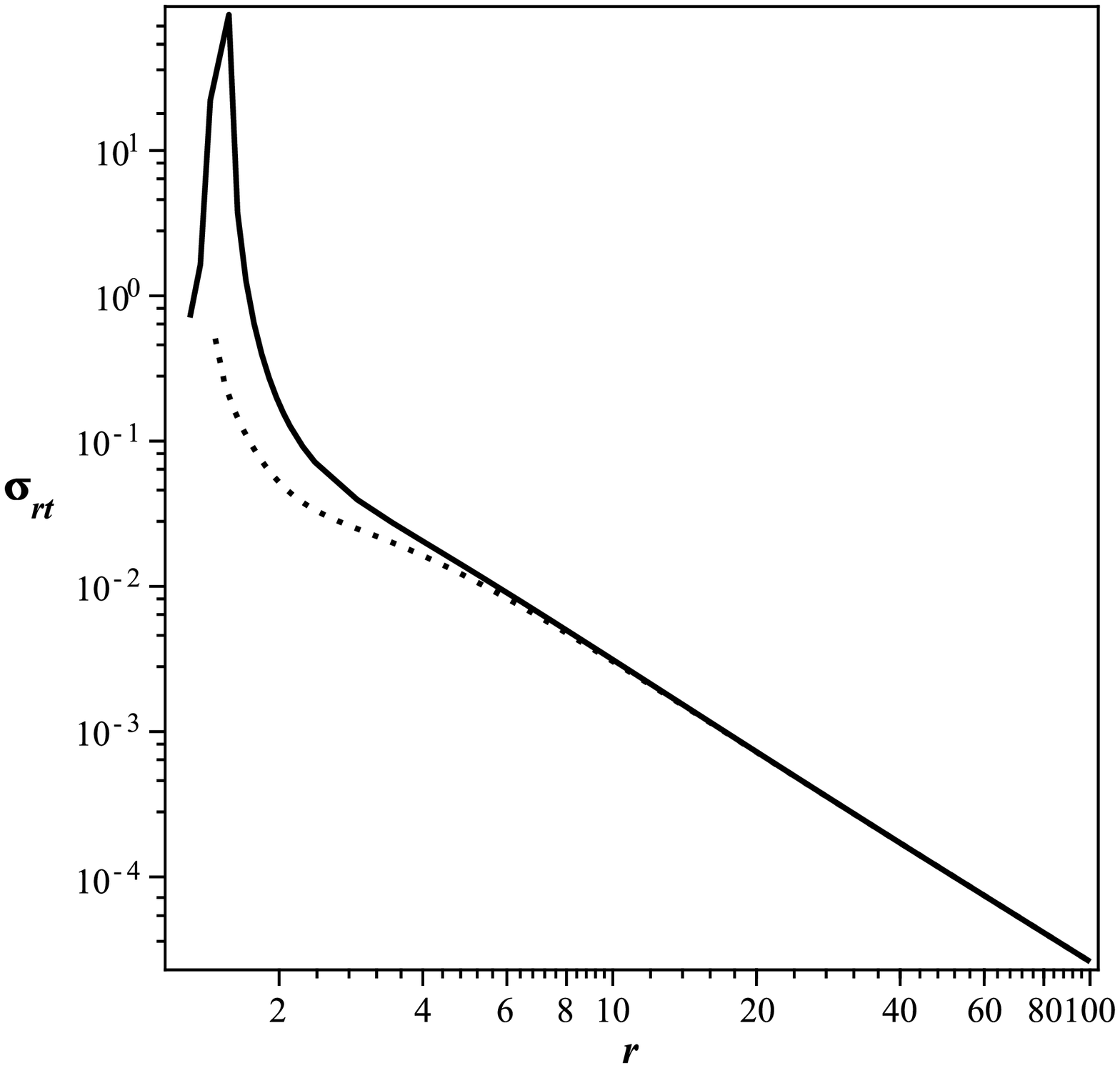}\epsfxsize=3.2in \epsfysize=2.3in \epsffile{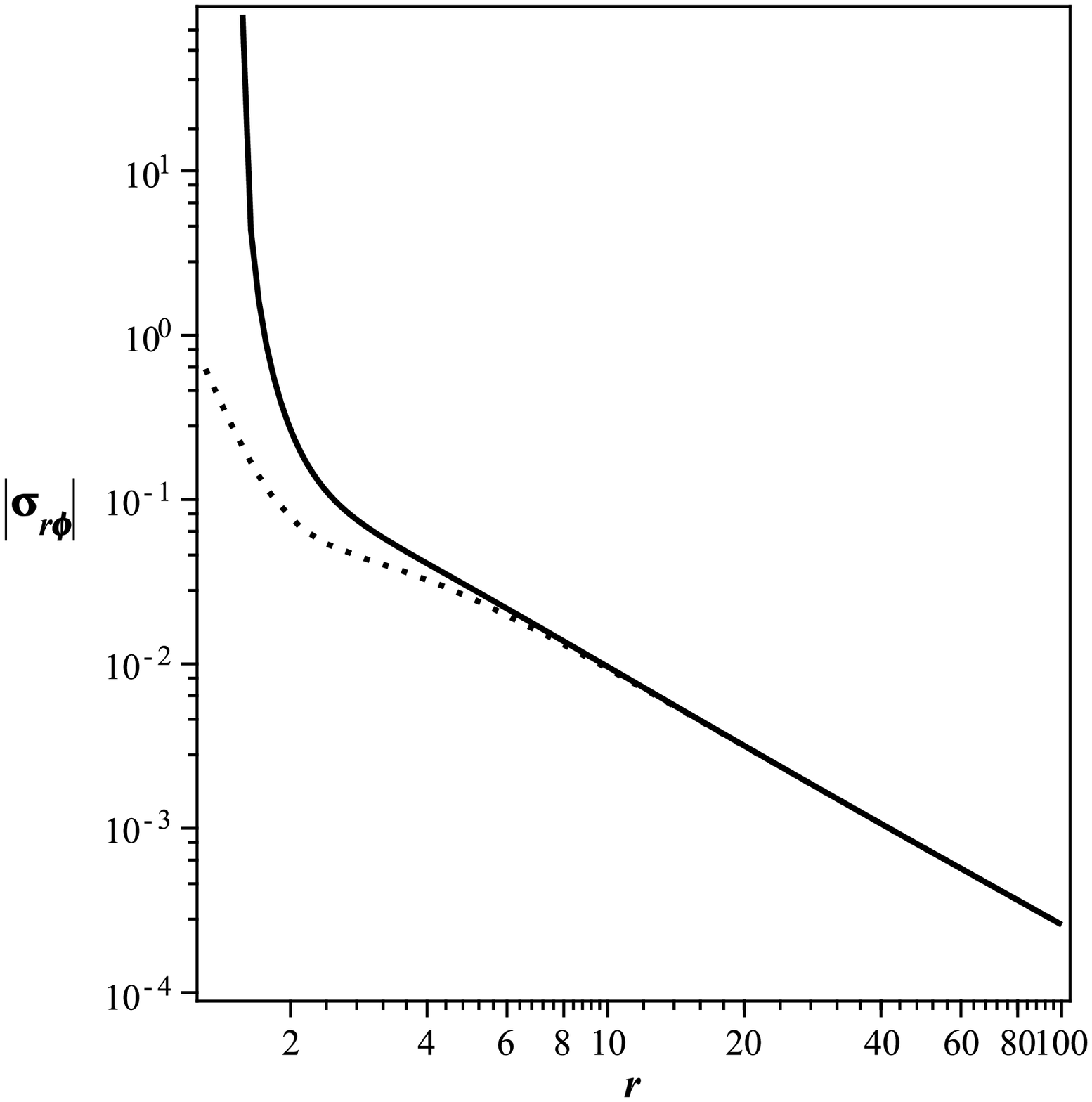}}
 \caption{\small{Influence of this simplifying in shear tensor components for $\Omega^{+}$ and $a=.9$. Solid curves are without that simplifying and dotted curve are with this simplifying}}
 \label{figure1}
\end{figure*}
\end{center}
Setting $l=\Omega r^2$ (Takahashi 2007a), the angular momentum can
be calculated. From equations (\ref{34}) and (\ref{14}), $\rho$ can
be derived as:
\begin{equation}\label{35}
\rho=\frac{j-l}{8\pi H_{\theta}r^2\nu\sigma^{\pm r}_{\phi}},
\end{equation}
where $\nu$ is (Takahashi 2007a):
\begin{equation}\label{35*}
\nu=\alpha a_{s}^{2}/\Omega_{k},
\end{equation}
which is the usual $\alpha$ prescription for viscosity(Shakura \&
Sunyaev 1973). In some papers $H_{\theta}$ is calculated by studying
the vertical structure such as Riffert \& Herold (1995), Abramowicz
et al.(1997) and Takahashi(2007b). In these papers $H_{\theta}$ was
calculated by vertical averaging, then introduced by other variable
such as $l$, $u_{t}$, $p$, $\rho$ and etc. If we use each of those
relations of $H_{\theta}$, we must solve all equations numerically.
But we want to derive an analytical solution therefore similar to
Takahashi(2007a) we use a simple form of $H_{\theta}$ as:
\begin{equation}\label{9}
H_{\theta}=\sqrt{\frac{5}{2}}\frac{a_{s}}{\Omega_{k}},
\end{equation}
 definitely, $\rho$ is the same in LNRF and BLF and is shown in figure 2
for ($\Omega^{-}$) and ($\Omega^{+}$) by assuming $a_{s}=.1$.
\input{epsf}
\begin{center}
\epsfxsize=3.1in \epsfysize=2.1in
\begin{figure*}
\centerline{\epsffile{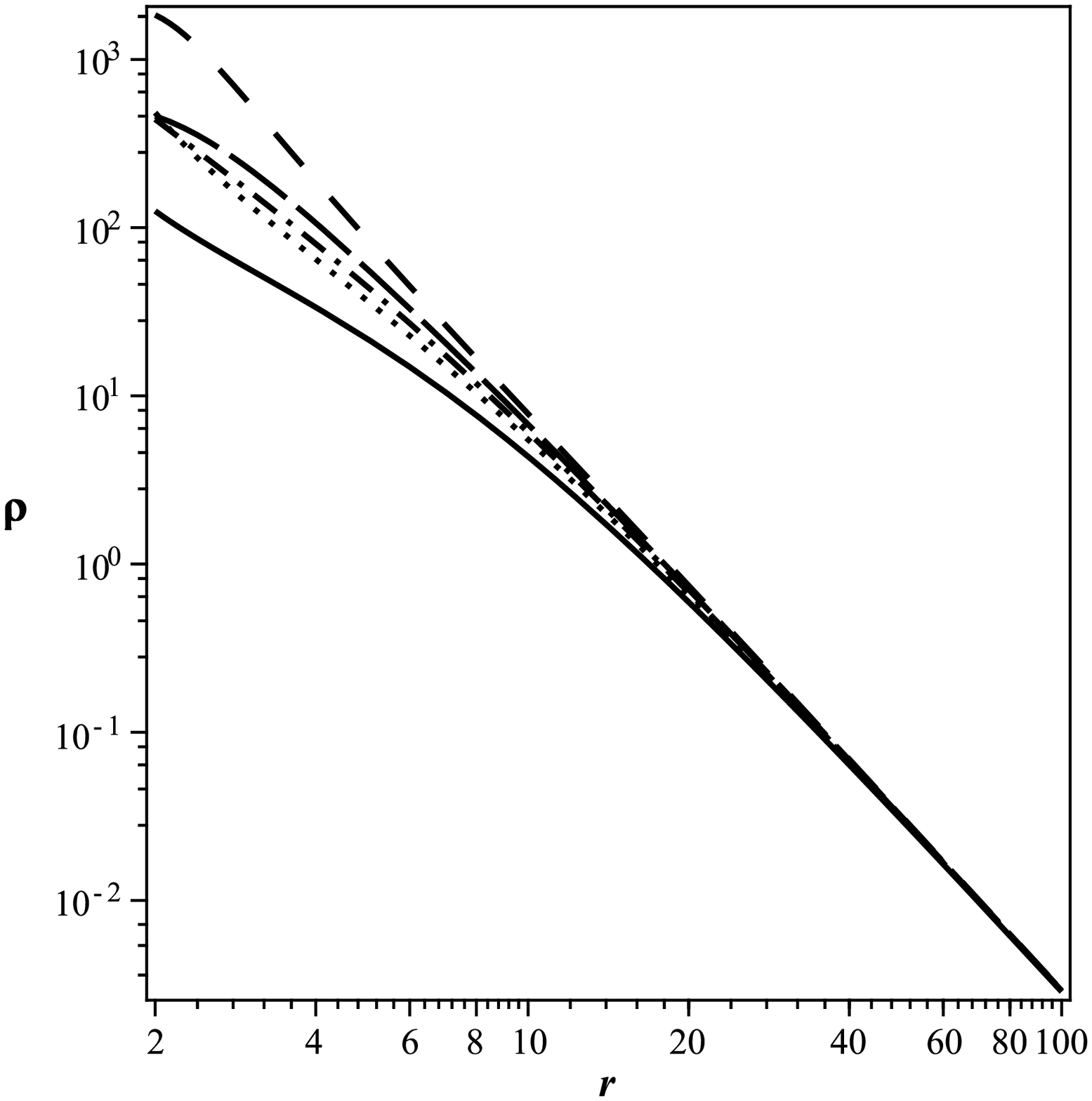}\epsfxsize=3.1in \epsfysize=2.1in
\epsffile{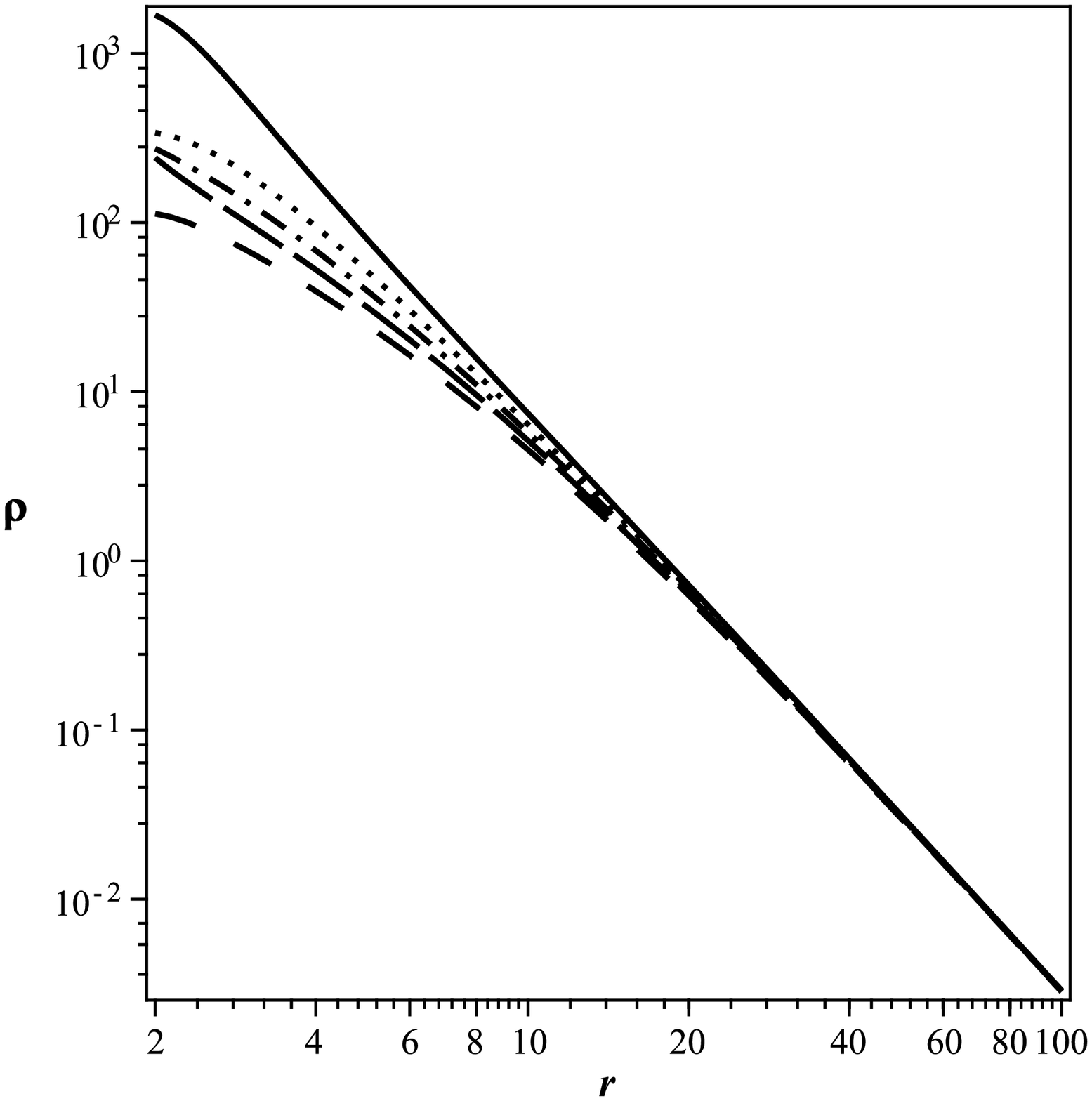}} \caption{\small{Surface density, left:
$\Omega^{-}$, right: $\Omega^{+}$ ($\alpha=.01$, $j=3$, solid:
$a=.9$, dotted: $a=.4$, dash-dotted: $a=0$, long-dashed: $a=-.4$ and
dash-spaced: $a=-.9$)}}
 \label{figure2}
\end{figure*}
\end{center}
Also $u^{\hat{r}}$ can be calculated from equation (\ref{11}) and $u_{\hat{t}}$ can
be found from equation (\ref{16}). Assuming $\dot{E}\simeq\dot{M}=1$ we have:
\begin{equation}\label{35'}
u^{\hat{r}}=\frac{1}{4\pi H_{\theta}r^2\rho},\qquad u_{\hat{t}}=-\dot{E}-\frac{t^{\hat{r}}_{\hat{t}}}{\rho u^{\hat{r}}}.
\end{equation}
We can calculate four velocity in BLF using the
transformation equations in Appendix 2, as:
\begin{equation}\label{36}
u_{t}=e_{t}^{\hat{\mu}}u_{\hat{\mu}}=\sqrt{\frac{\Delta\Sigma}{A}}u_{\hat{t}}-\frac{2ar}{\sqrt{A\Sigma}}u_{\hat{\phi}},
\end{equation}
\begin{equation}\label{37}
u_{r}=e_{r}^{\hat{\mu}} u_{\hat{\mu}}=\sqrt{\frac{\Sigma}{\Delta}} u_{\hat{r}},
\end{equation}
\begin{equation}\label{38}
u_{\phi}=e_{\phi}^{\hat{\mu}}u_{\hat{\mu}}=\sqrt{\frac{A}{\Sigma}}u_{\hat{\phi}}.
\end{equation}
Now, we may see ($\left|u_{t}\right|,\left|u_{\hat{t}}\right|$),
($\left|u_{r}\right|$, $\left|u_{\hat{r}}\right|$) and
($\left|u_{\phi}\right|,\left|u_{\hat{\phi}}\right|$)in figure 3.
\input{epsf}
\begin{center}
\epsfxsize=3.1in \epsfysize=2.1in
\begin{figure*}
\centerline{\epsffile{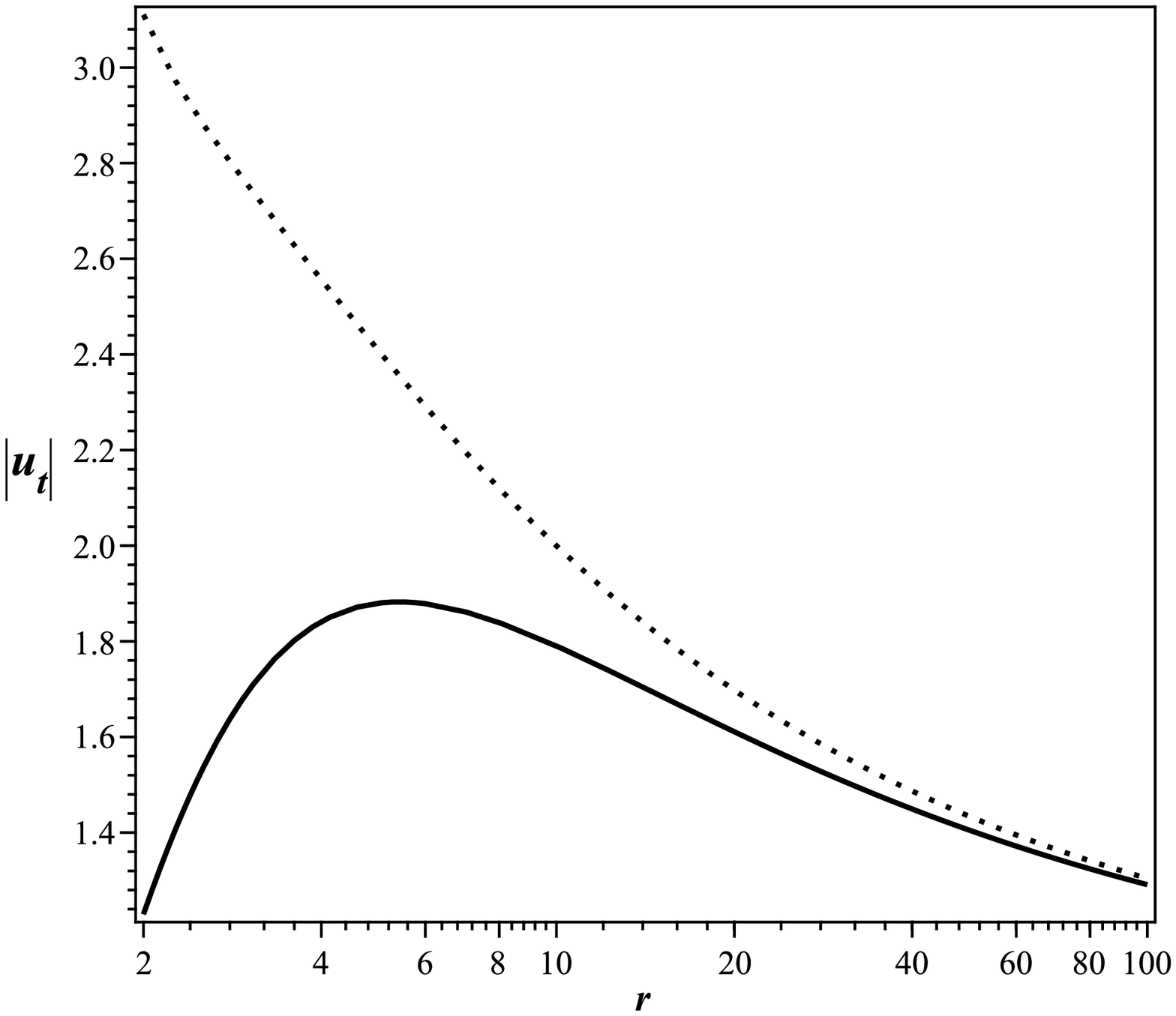} \epsfxsize=3.1in \epsfysize=2.1in
\epsffile{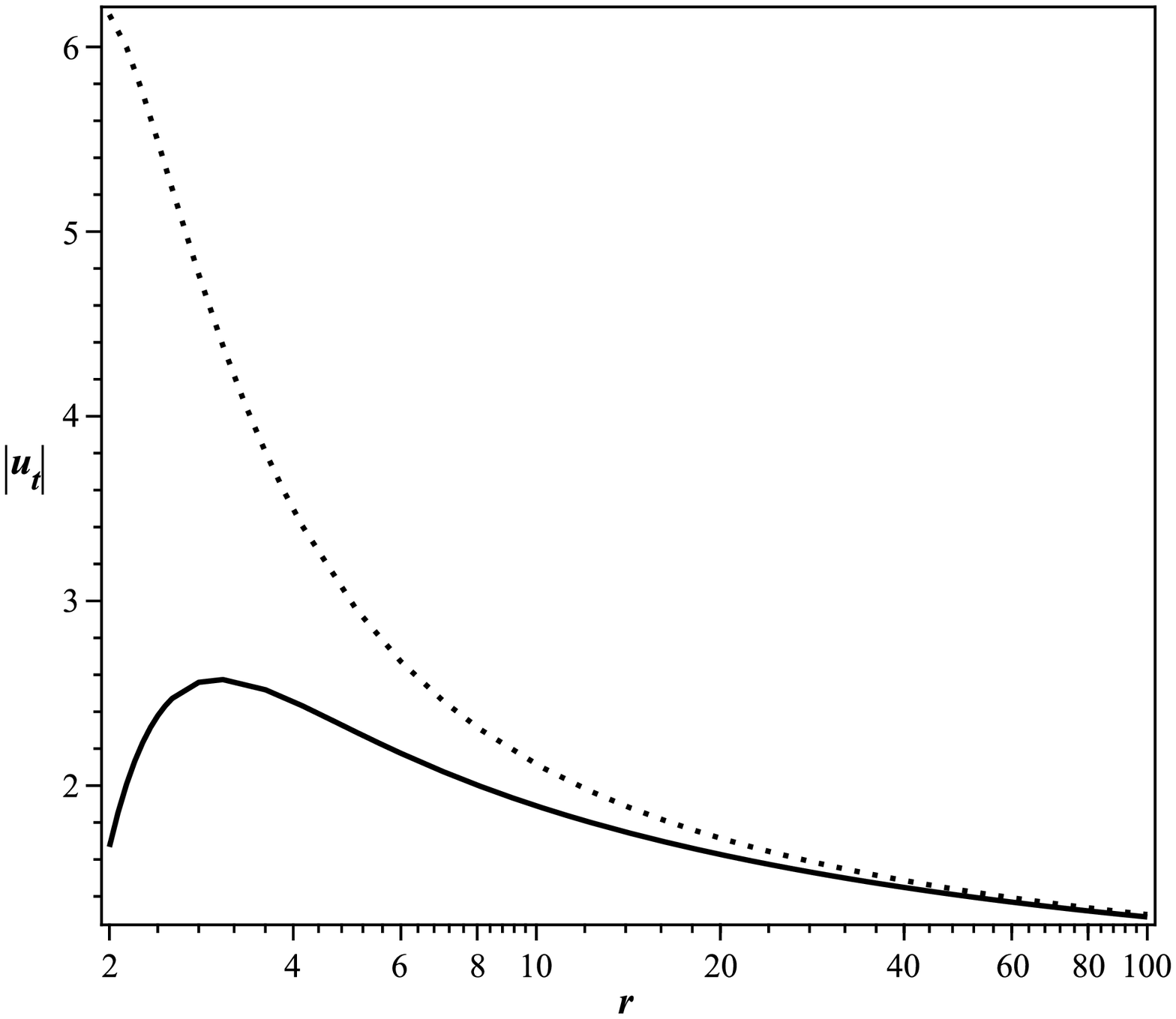}} \epsfxsize=3.1in \epsfysize=2.1in
\centerline{\epsffile{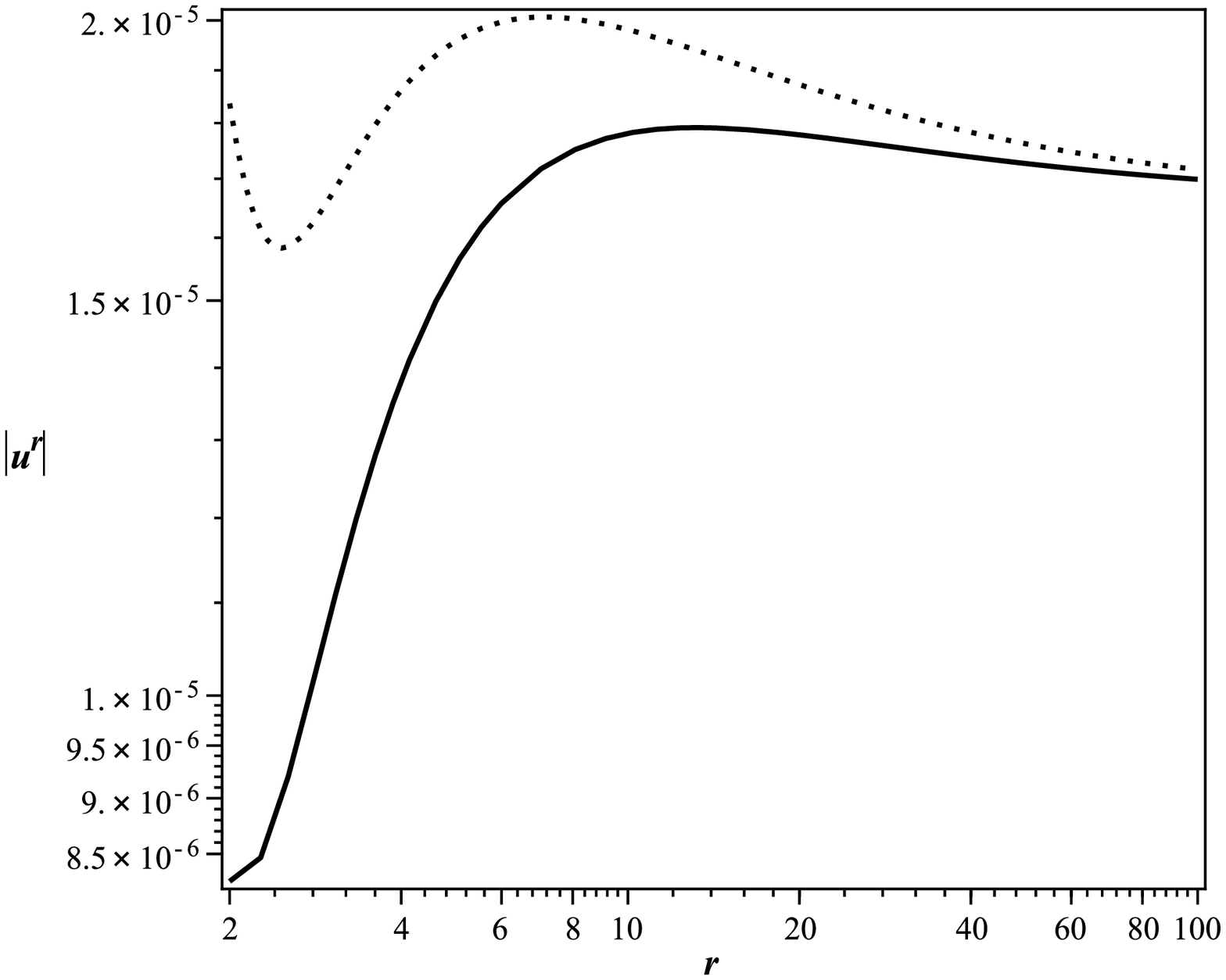}\epsfxsize=3.1in \epsfysize=2.1in
\epsffile{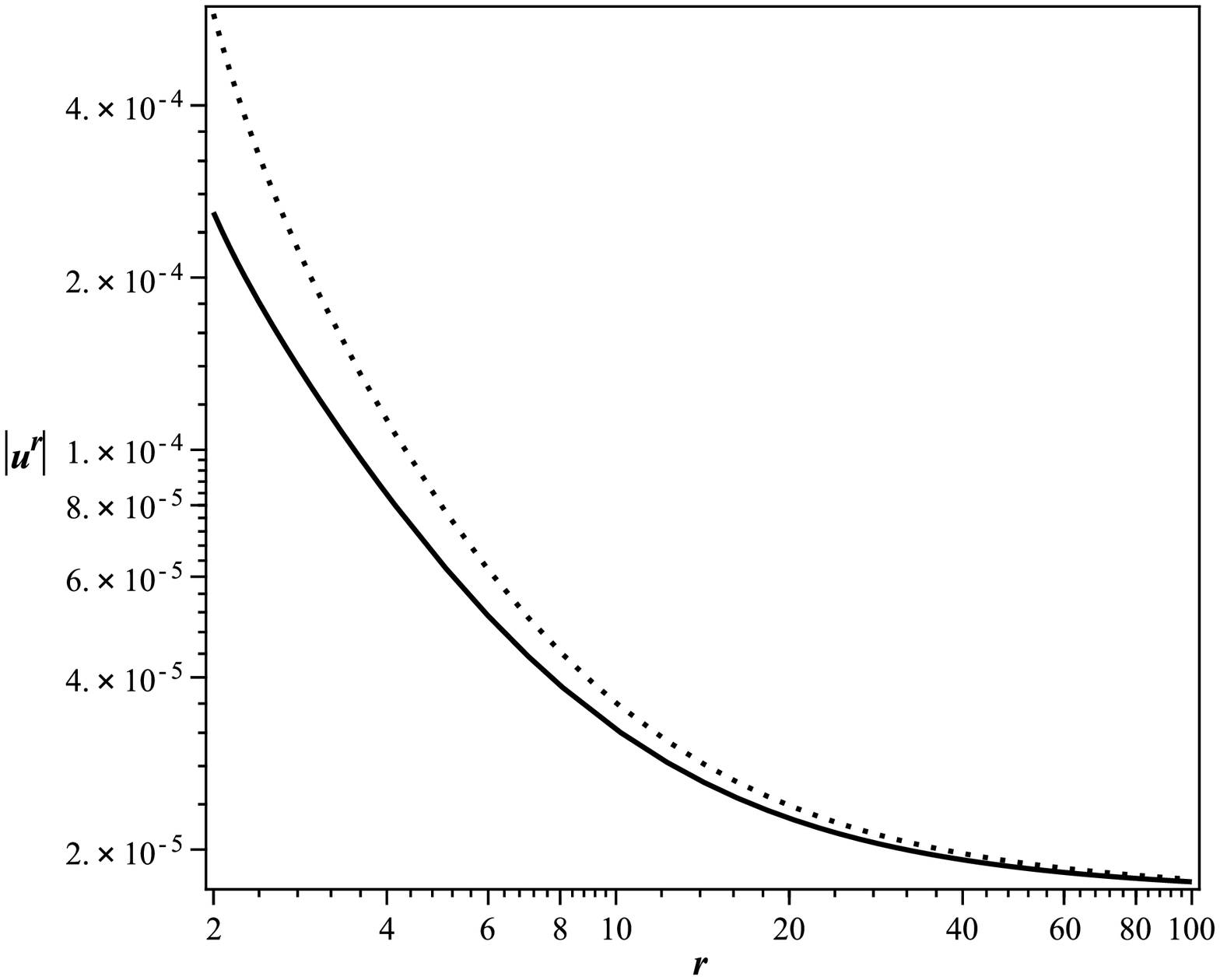}} \epsfxsize=3.1in \epsfysize=2.1in
\centerline{\epsffile{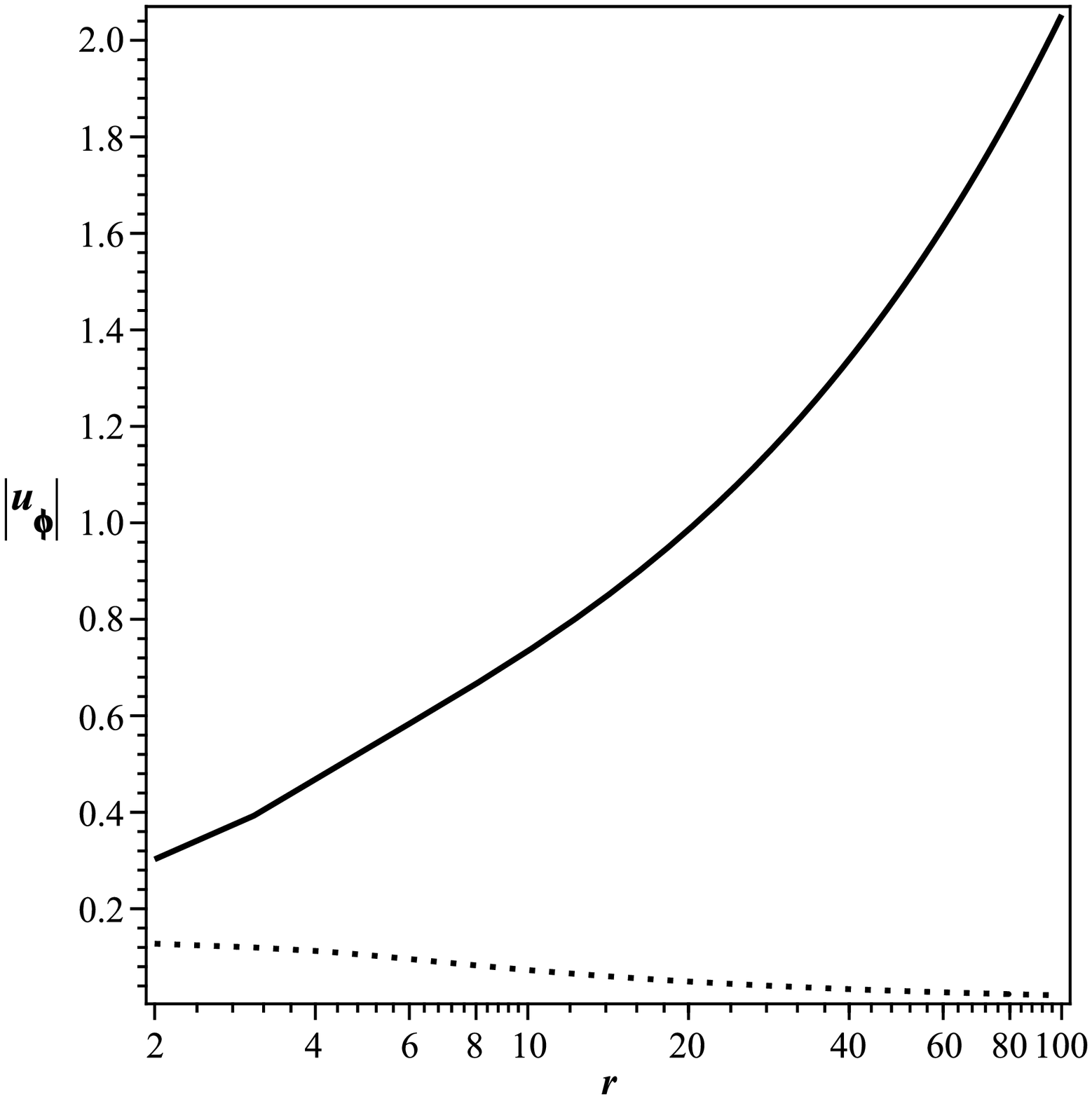},\epsfxsize=3.1in
\epsfysize=2.1in \epsffile{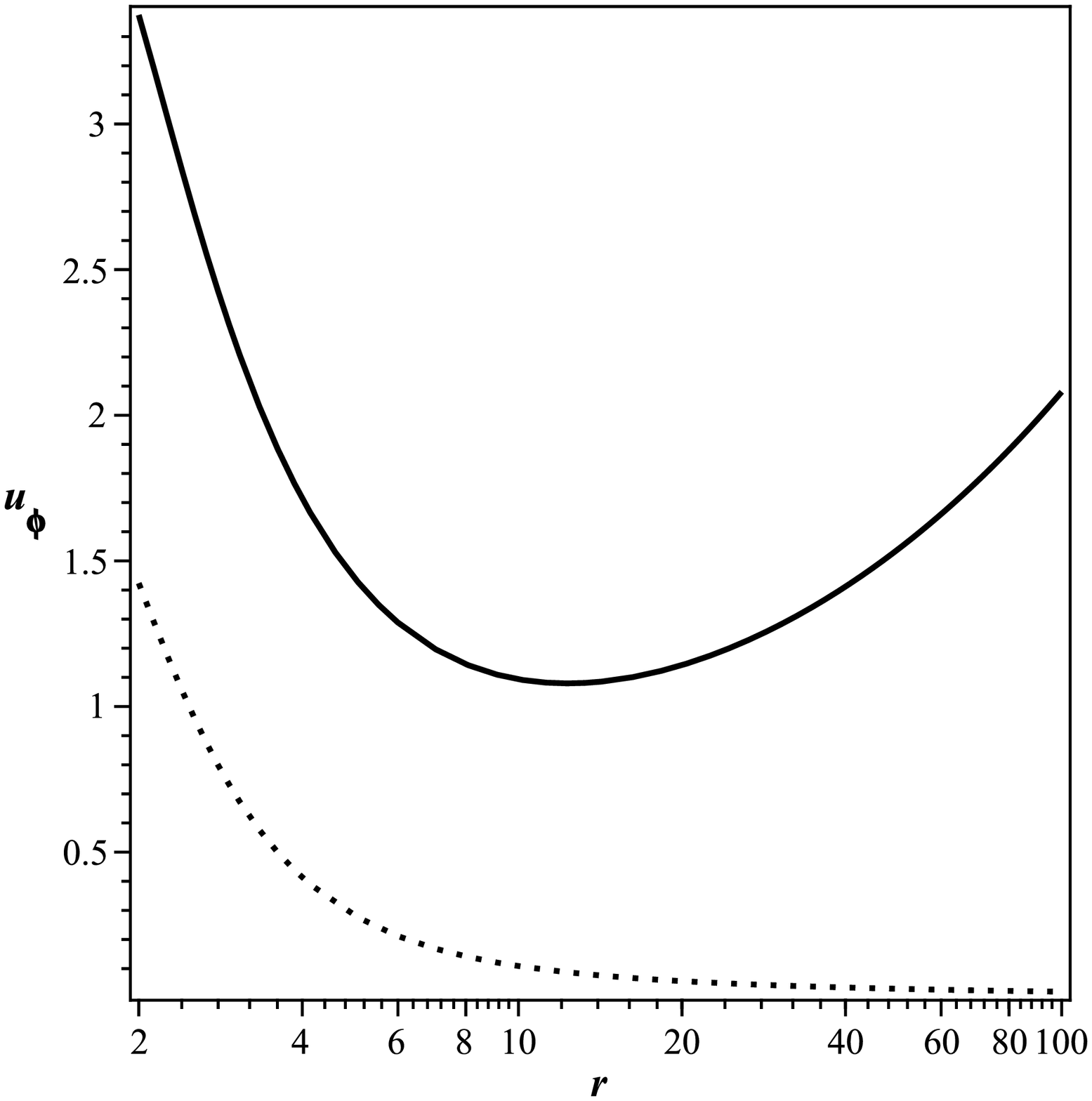}} \caption{\small{Four
velocity in LNRF and BLF. Left: $\Omega^{-}$, right: $\Omega^{+}$
      ($\alpha=.01$, $j=3$, $a=-.9$ and $a_{s}=.1$, solid: in BLF and dotted: in LNRF)}}
\label{figure3}
\end{figure*}
\end{center}
Four velocity and $\left|\Omega\right|$ in BLF($\left|u_{\theta}=0\right|$) are
shown in figure 4 setting $a_{s}=.1$.
\input{epsf}
\begin{center}
\epsfxsize=3.1in \epsfysize=2.1in
\begin{figure*}
\centerline{\epsffile{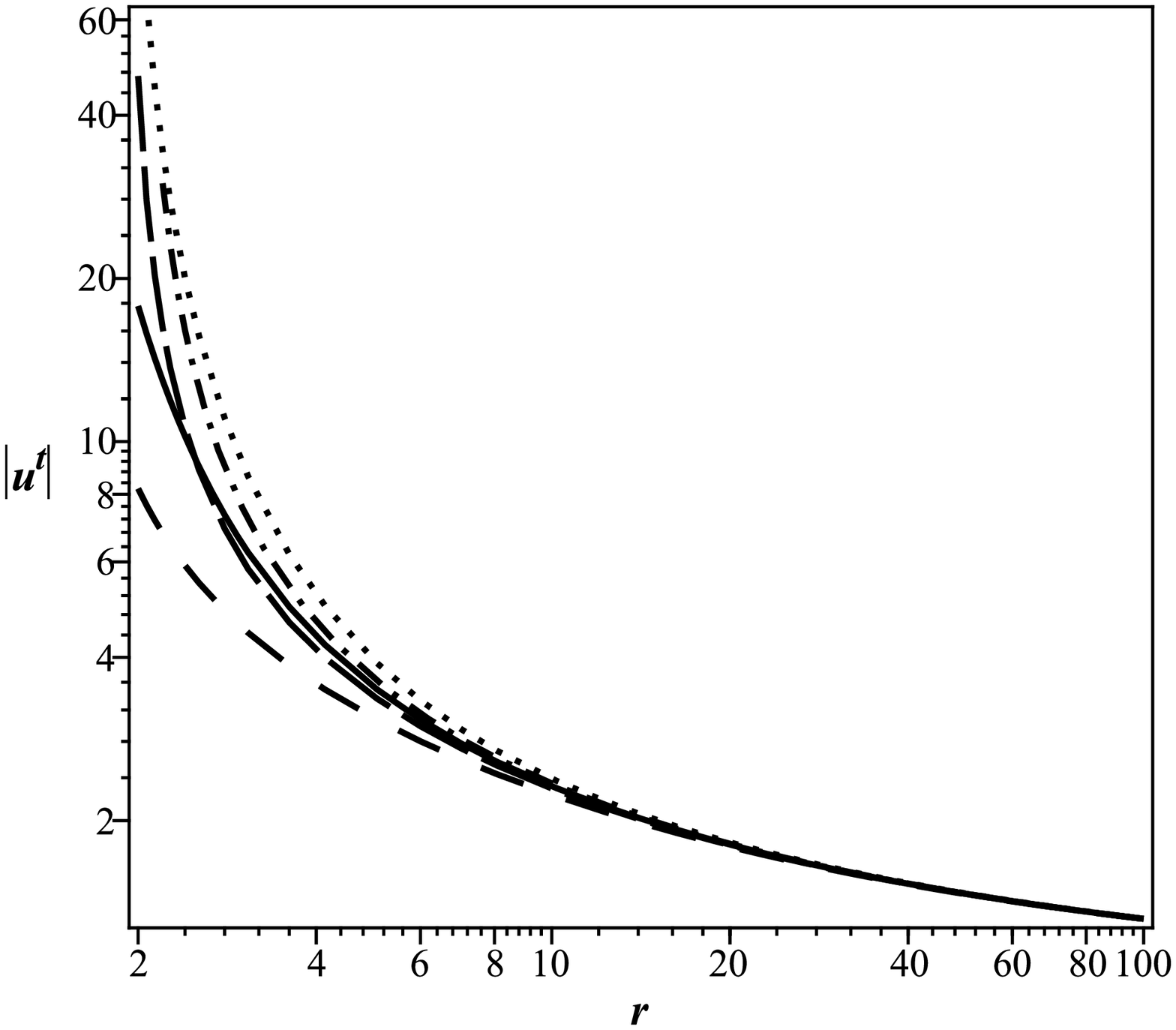} \epsfxsize=3.2in \epsfysize=2.1in
\epsffile{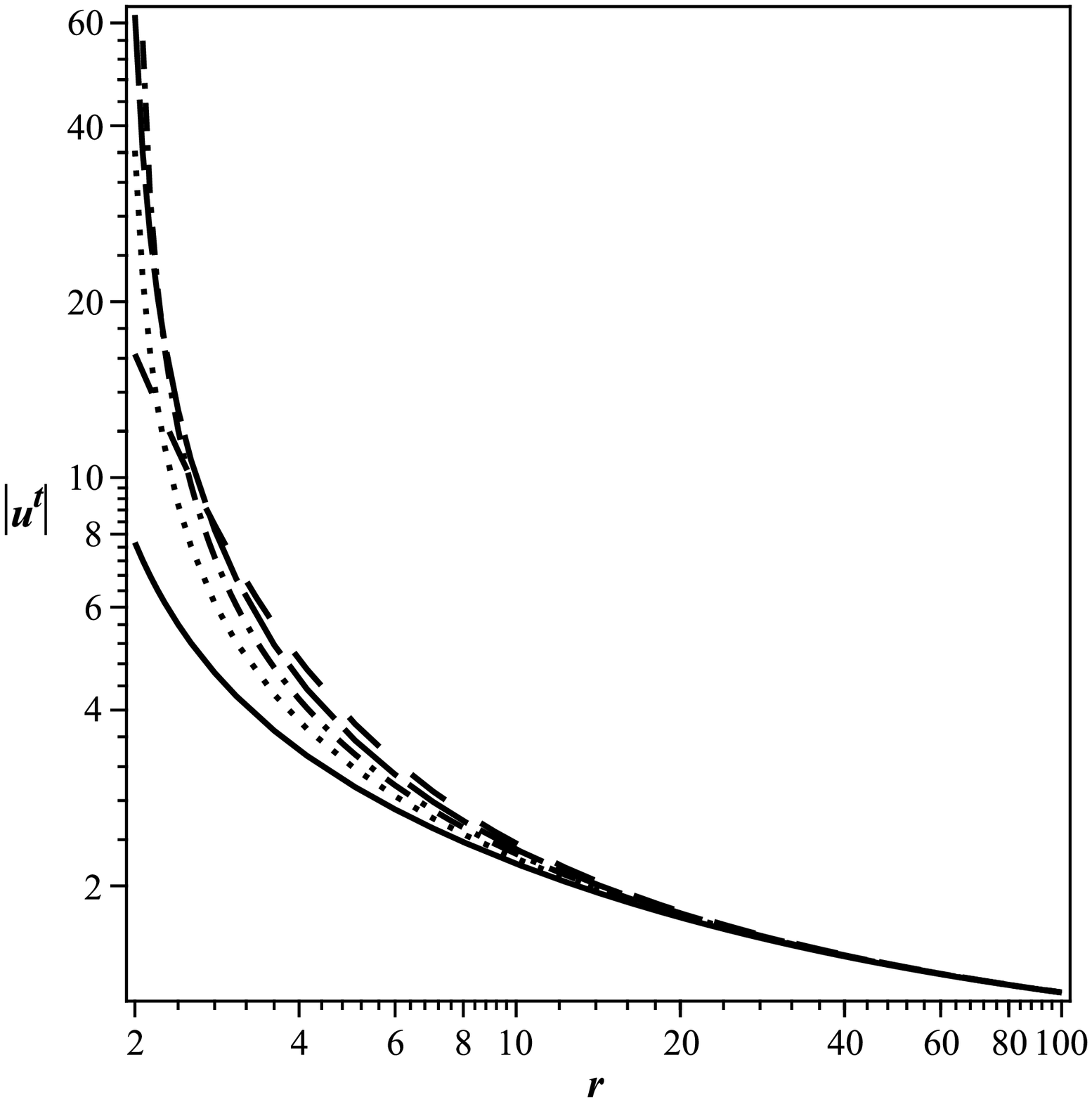}} \epsfxsize=3.2in \epsfysize=2.1in
\centerline{\epsffile{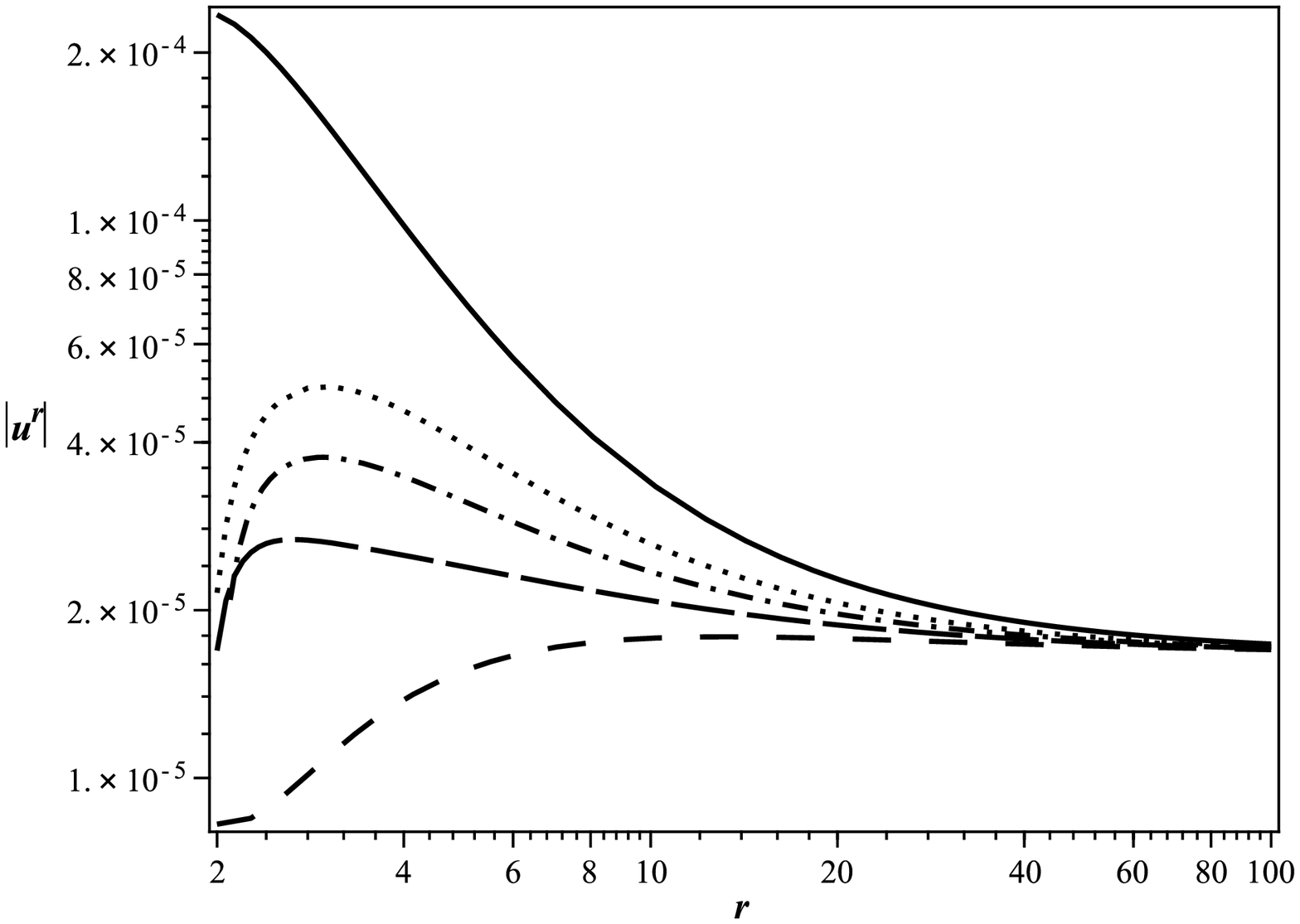}\epsfxsize=3.2in \epsfysize=2.1in
\epsffile{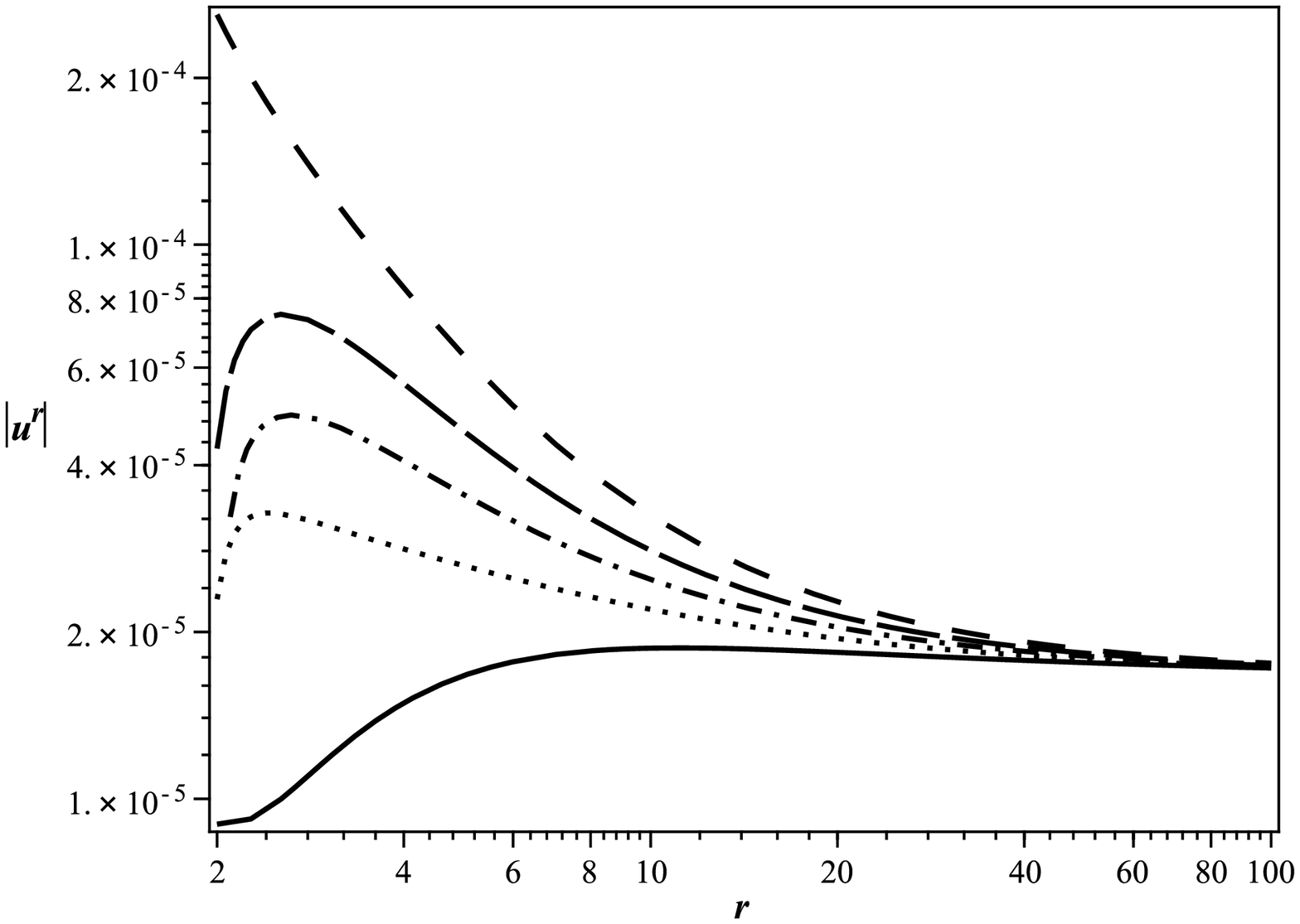}} \epsfxsize=3.2in \epsfysize=2.1in
\centerline{\epsffile{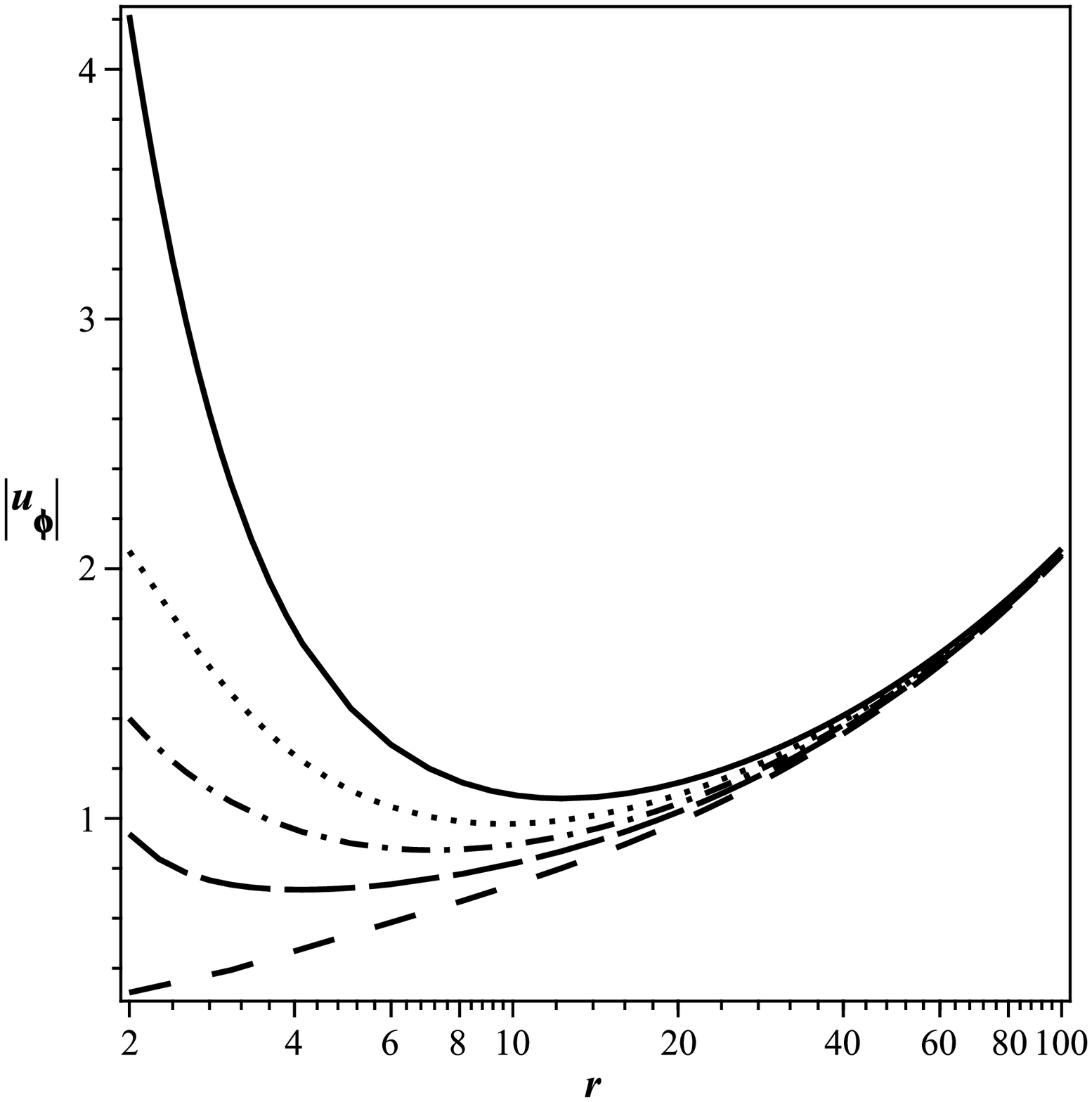}\epsfxsize=3.2in \epsfysize=2.1in
\epsffile{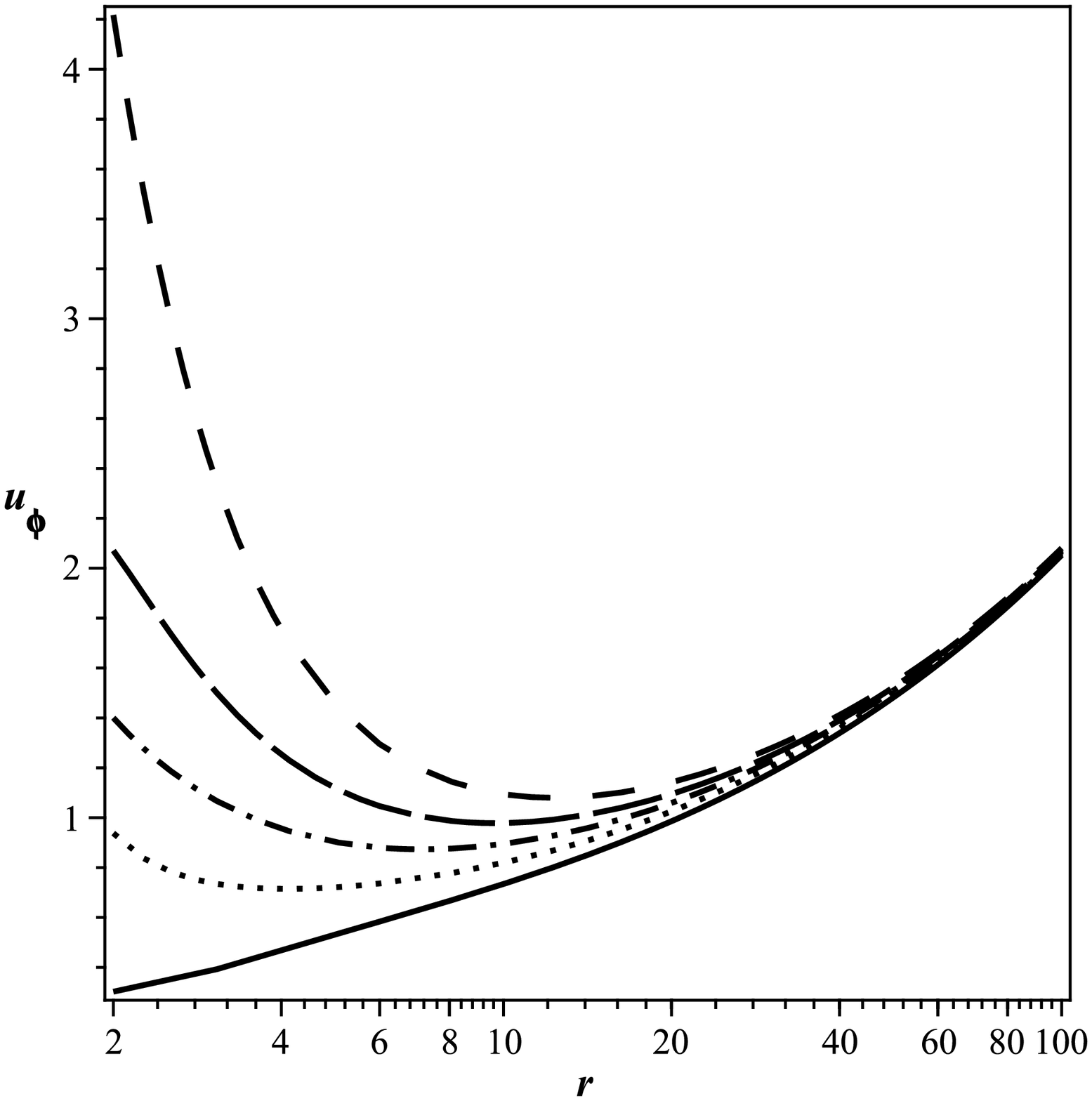}} \epsfxsize=3.2in \epsfysize=2.1in
\centerline{\epsffile{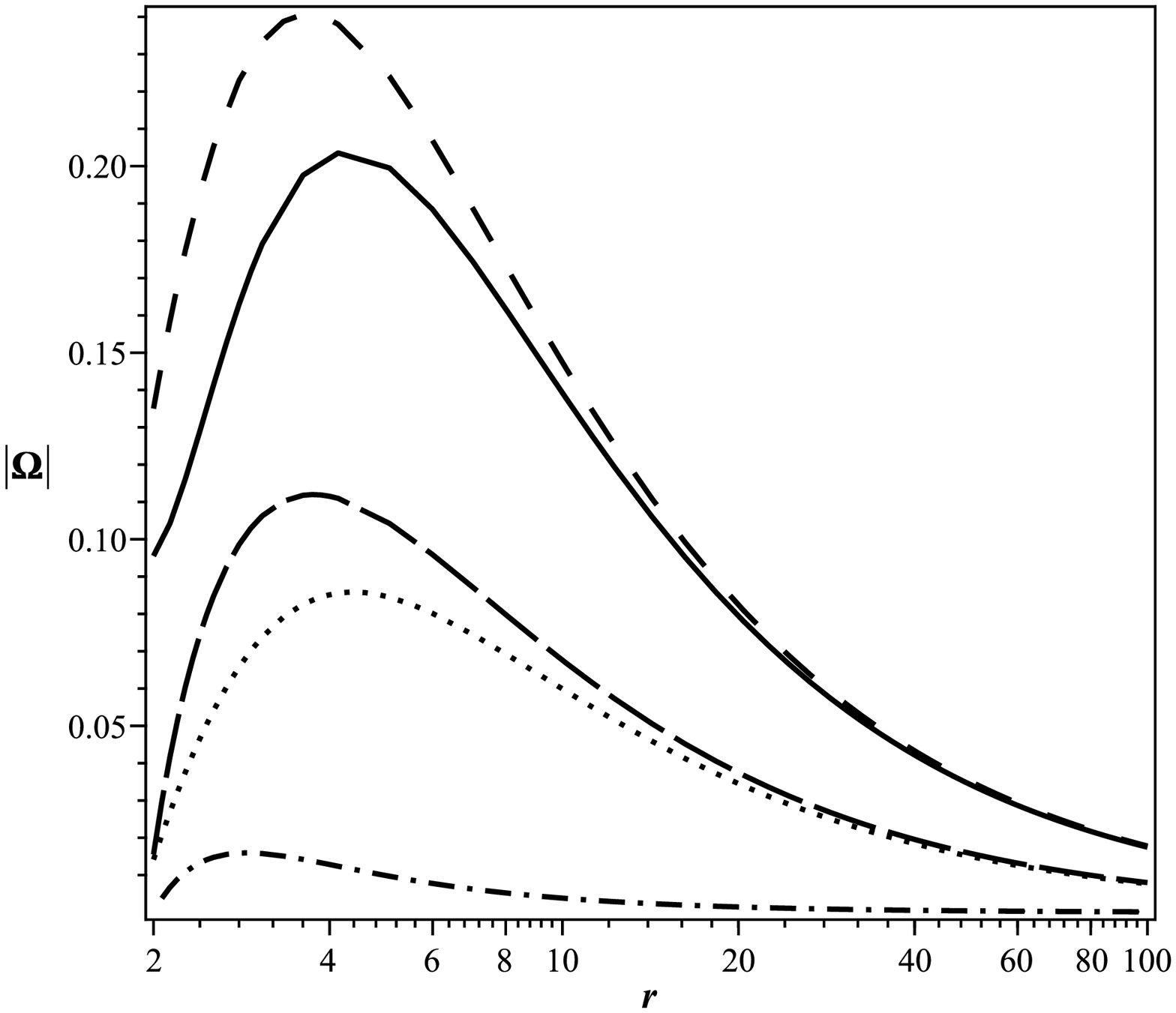}\epsfxsize=3.2in \epsfysize=2.1in
\epsffile{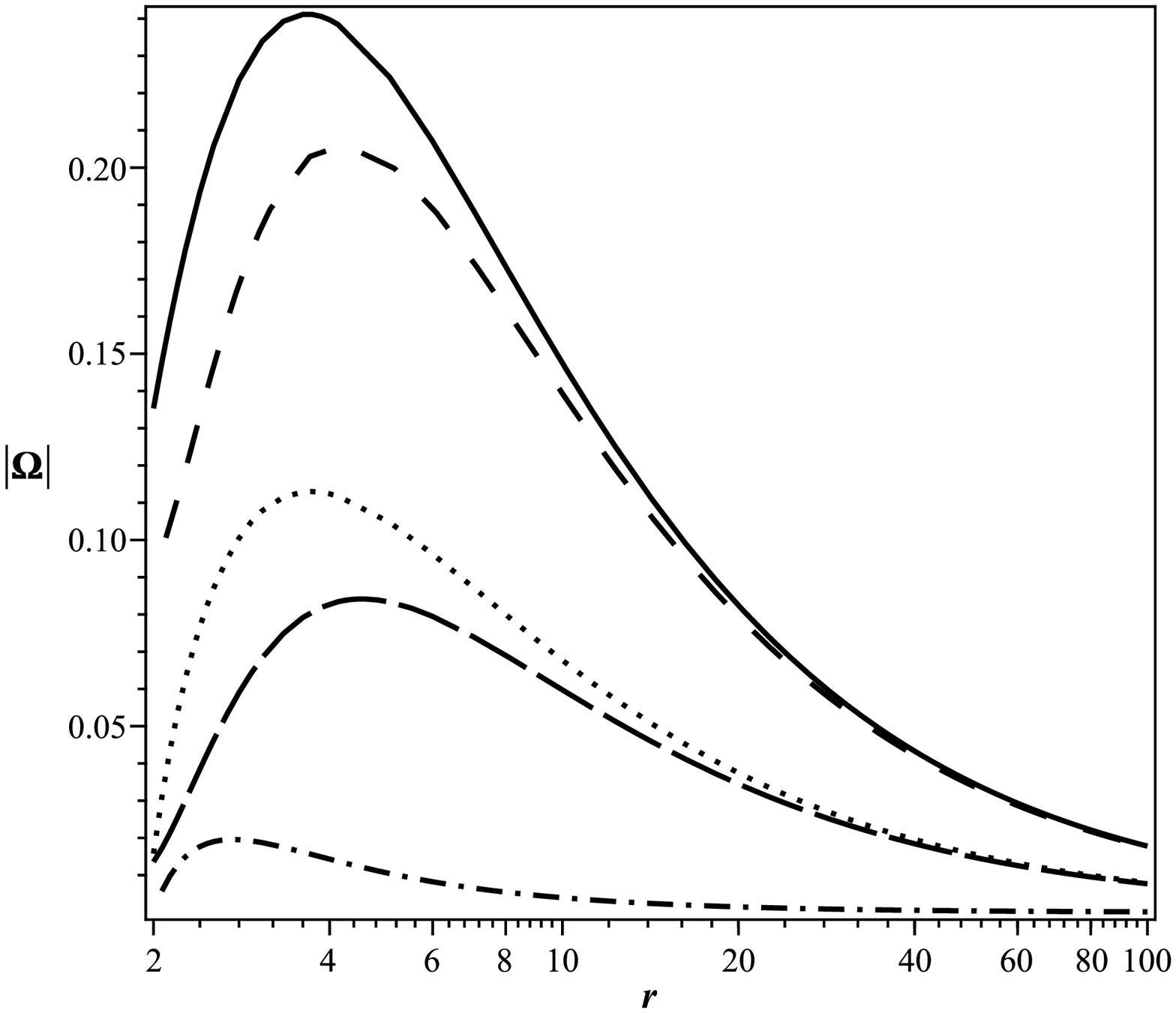}} \caption{\small{Four velocity and
$\left|\Omega\right|$ in the BLF. Left: $\Omega^{-}$, right:
$\Omega^{+}$($\alpha=.01$, $j=3$, solid: $a=.9$, dotted: $a=.4$,
dash-dotted: $a=0$, long-dashed: $a=-.4$ and dash-spaced: $a=-.9$)}}
\label{figure4}
\end{figure*}
\end{center}
In equation (\ref{35*}), $0<\alpha<1$ is viscosity coefficient in
$\alpha$ prescription which influences on redistribution of
energy and momentum. The influence of these parameters are shown in
figure 5 with $a_{s}=.1$
\input{epsf}
\begin{center}
\epsfxsize=3.2in \epsfysize=2.3in
\begin{figure*}
\centerline{\epsffile{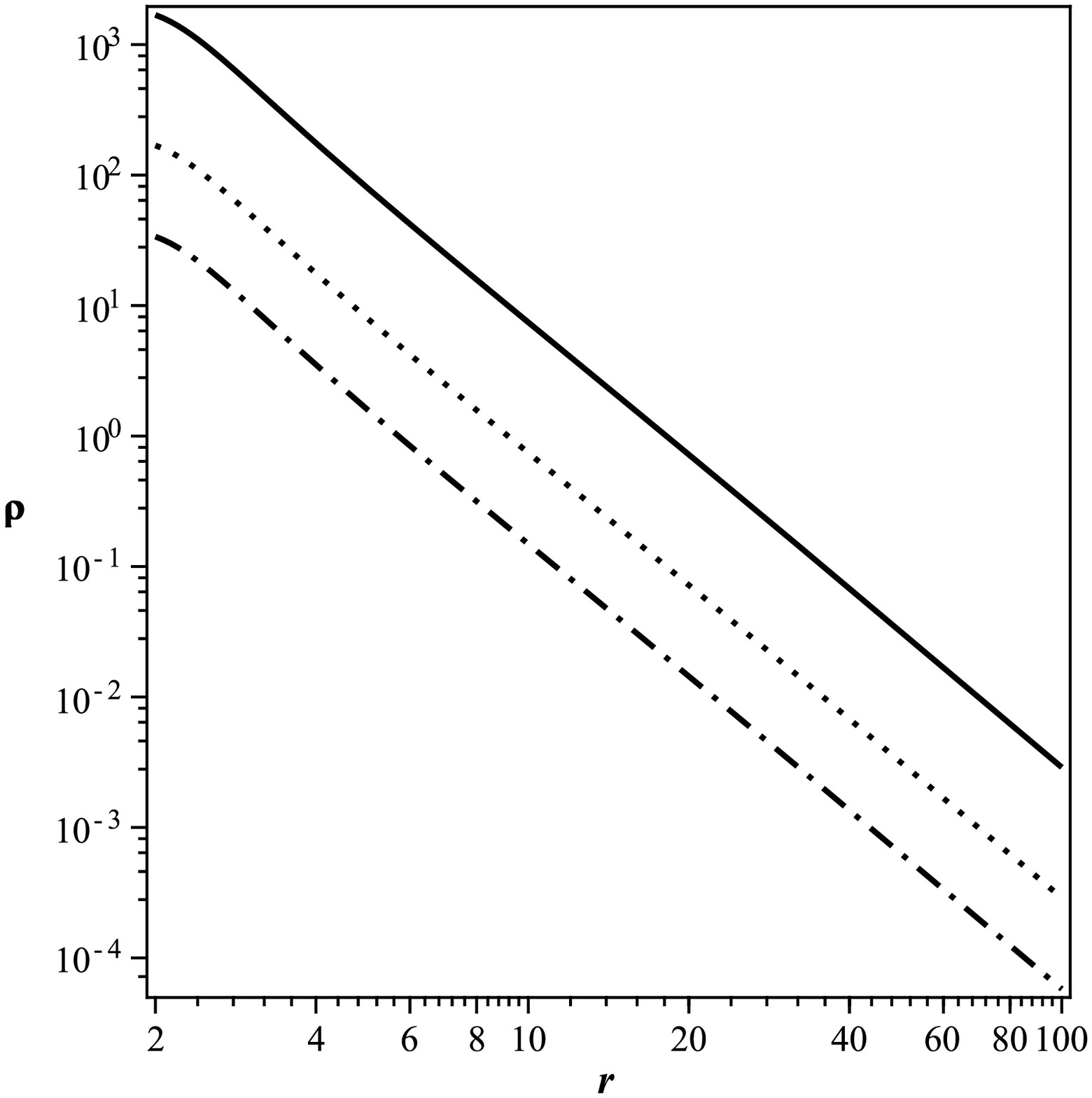} \epsfxsize=3.2in
\epsfysize=2.3in \epsffile{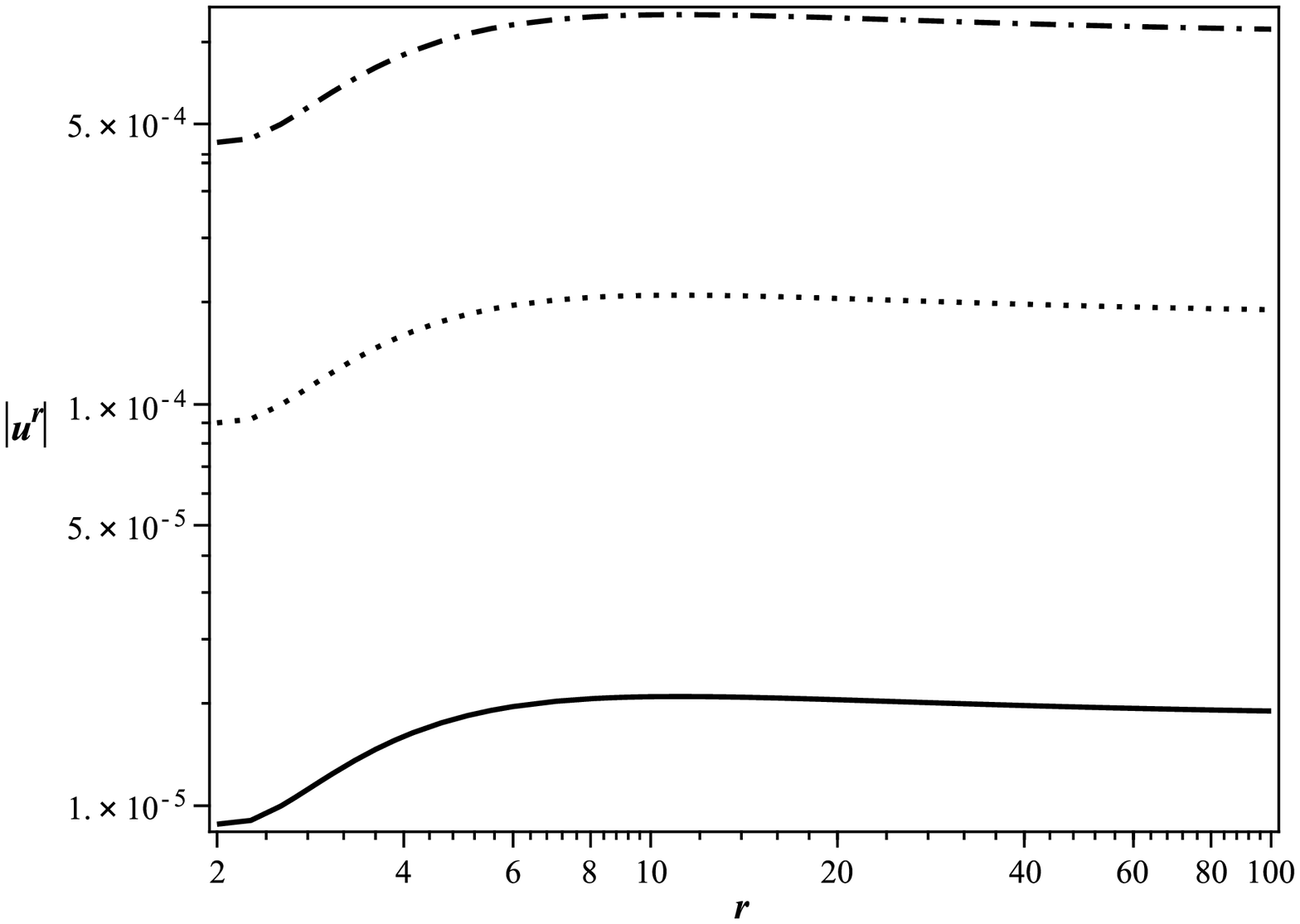}} \caption{\small{Influence
of $\alpha$ on density and radial four velocity for
$\Omega^{+}$($a=.9$, $j=3$ and $a_{s}=.1$. Solid: $\alpha=.01$,
dotted: $\alpha=.1$ and dash-dotted: $\alpha=.5$)}} \label{figure 5}
\end{figure*}
\end{center}
The four velocity and density can be derived for various total inward
flux of angular momentum (for example $j=1,2$ and 3) which are
shown in figure 6.

\input{epsf}
\begin{center}
\epsfxsize=3.2in \epsfysize=2.3in
\begin{figure*}
\centerline{\epsffile{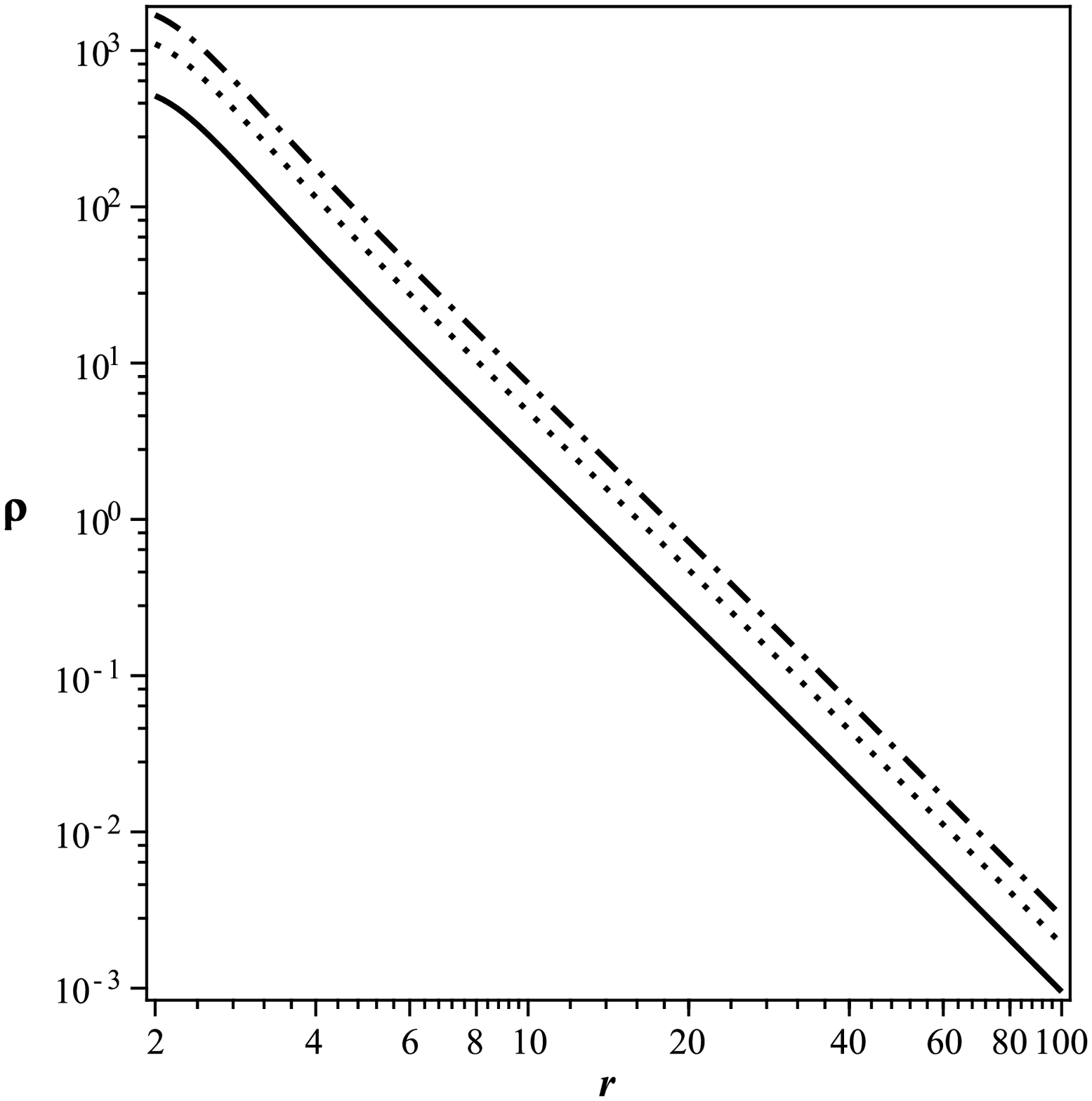} \epsfxsize=3.2in \epsfysize=2.3in
\epsffile{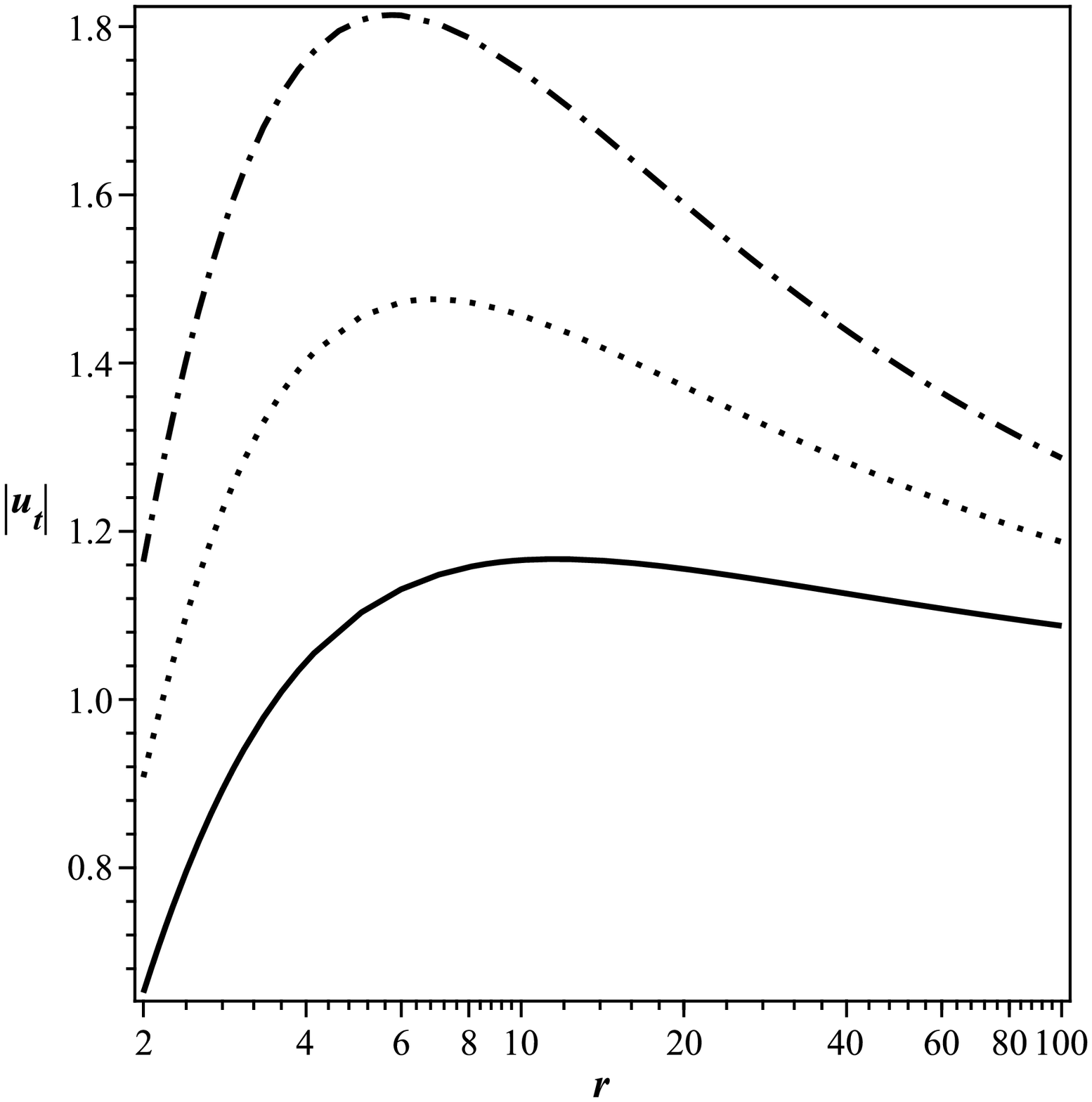}} \epsfxsize=3.2in \epsfysize=2.3in
\centerline{\epsffile{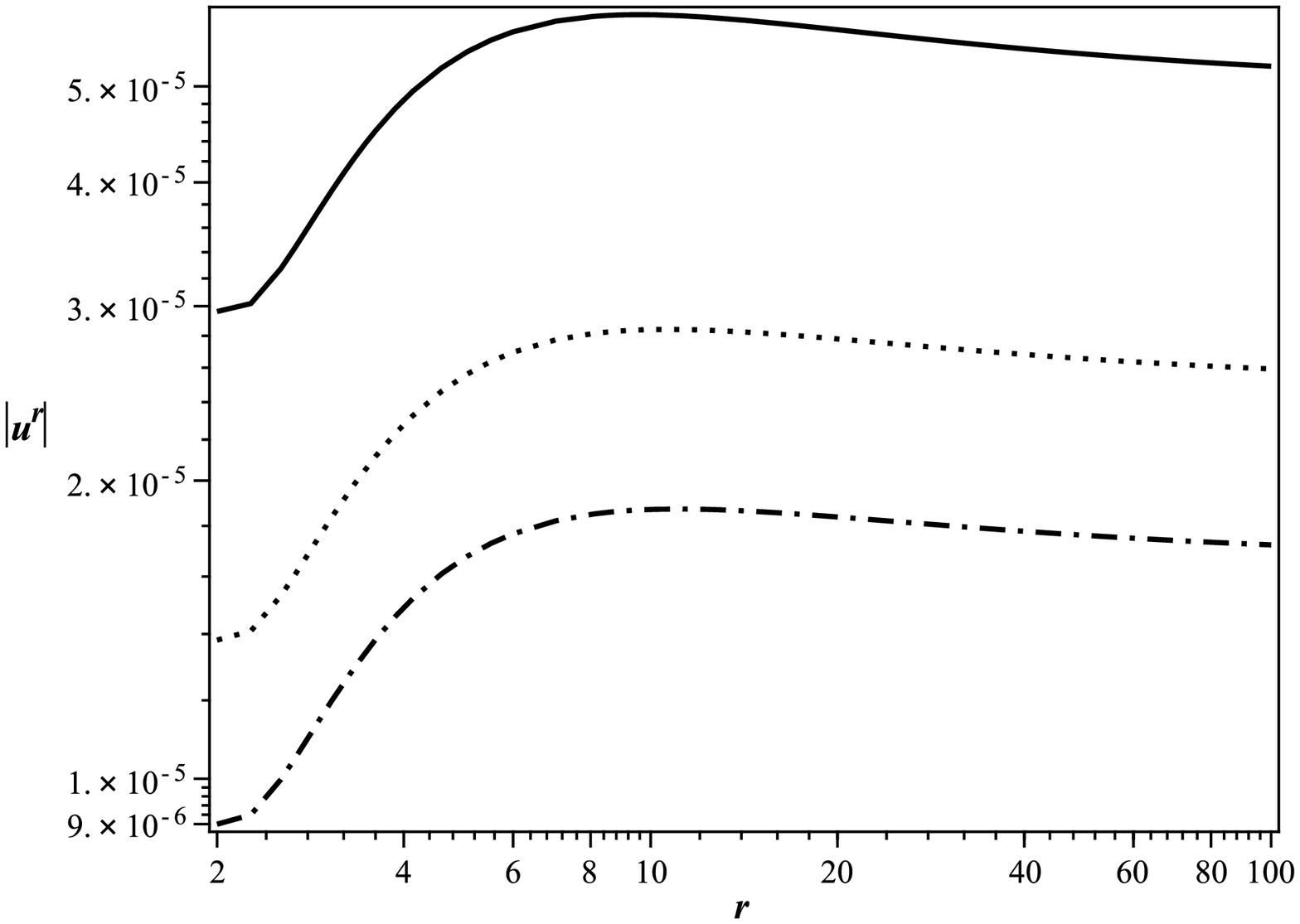}\epsfxsize=3.2in \epsfysize=2.3in
\epsffile{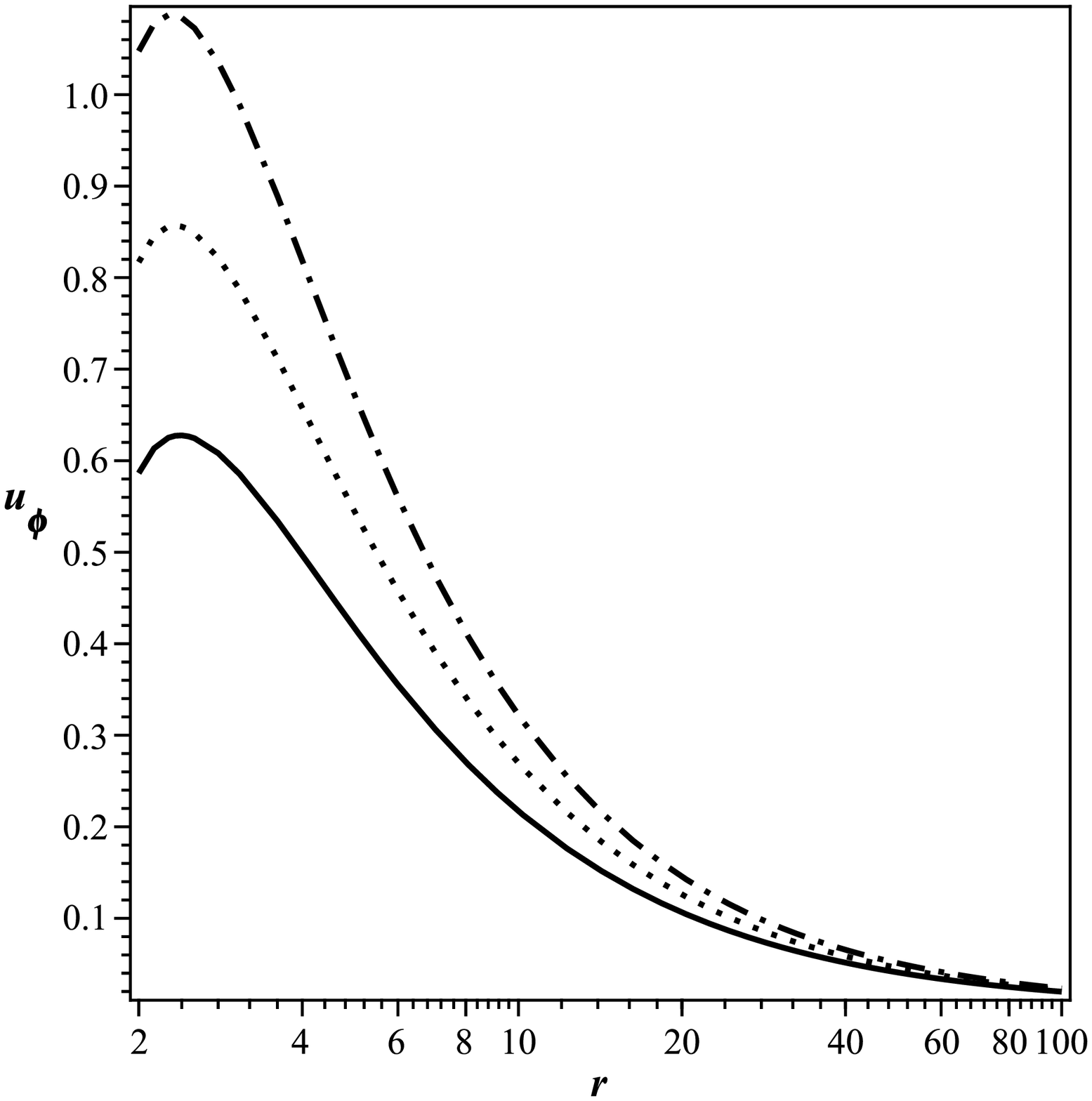}} \epsfxsize=3.2in \epsfysize=2.3in
\caption{\small{Influence of j in density and the four velocity of
$\Omega^{+}$ in BLF($\alpha=.01$, $a=.9$ and $a_{s}=.1$. Solid:
$j=1$, dotted: $j=2$ and dash-spaced: $j=3$)}} \label{figure6}
\end{figure*}
\end{center}
\subsection{Influence of $r-t$ component on the four velocity}
In pervious papers such as Abramowicz et al. (1996 and 1997), Gammie
\& Popham (1998), Manmoto (2000) and Takahashi (2007a,b), only the
$r-\phi$ component of shear tensor was assumed to be nonzero in FRF,
but in section 4 we showed that the $r-t$ component is also nonzero.
The $r-t$ component of shear tensor in LNRF results from covariant
derivative of $u_{\hat{t}}$. In general relativity the gravitating
field causes time dilation. Due to this time dilation, the
coordinate time(t) and proper time($\tau$) are not the
same($dt>d\tau$), therefor $u^{\hat{t}}$ in LNRF can be derived as:
\begin{equation}\label{6}
u^{\hat{t}}=\frac{d\hat{t}}{d\tau}.
\end{equation}
$u^{\hat{t}}=\left|u_{\hat{t}}\right|$ is different in each radius
and also tangent of $u_{\hat{t}}$ creates the r-t component of shear
tensor. To see the influence of this component, we derive four
velocity with both $r-t$ component of shear tensor and without it.
If we set $\sigma^{r}_{t}=0 $ in the pervious section the figures
are similar to that of the previous papers. From equations
(\ref{35}) and (\ref{35'}) it is clear that $r-t$ component affects
$u_{t}$ in all frames because without this component, the second
term of $u_{t}$ in equation (\ref{35'}) vanishes. If we calculate
$u^{\mu}=g^{\mu\nu}u_{\nu}$, the $r-t$ component also affects
$u^{\phi}$ in BLF. The influences of $r-t$ component of shear tensor
are shown in figure 7, solid curves are with this component and
dotted curves are without it.
\input{epsf}
\begin{center}
\epsfxsize=3.2in \epsfysize=2.3in
\begin{figure*}
\centerline{\epsffile{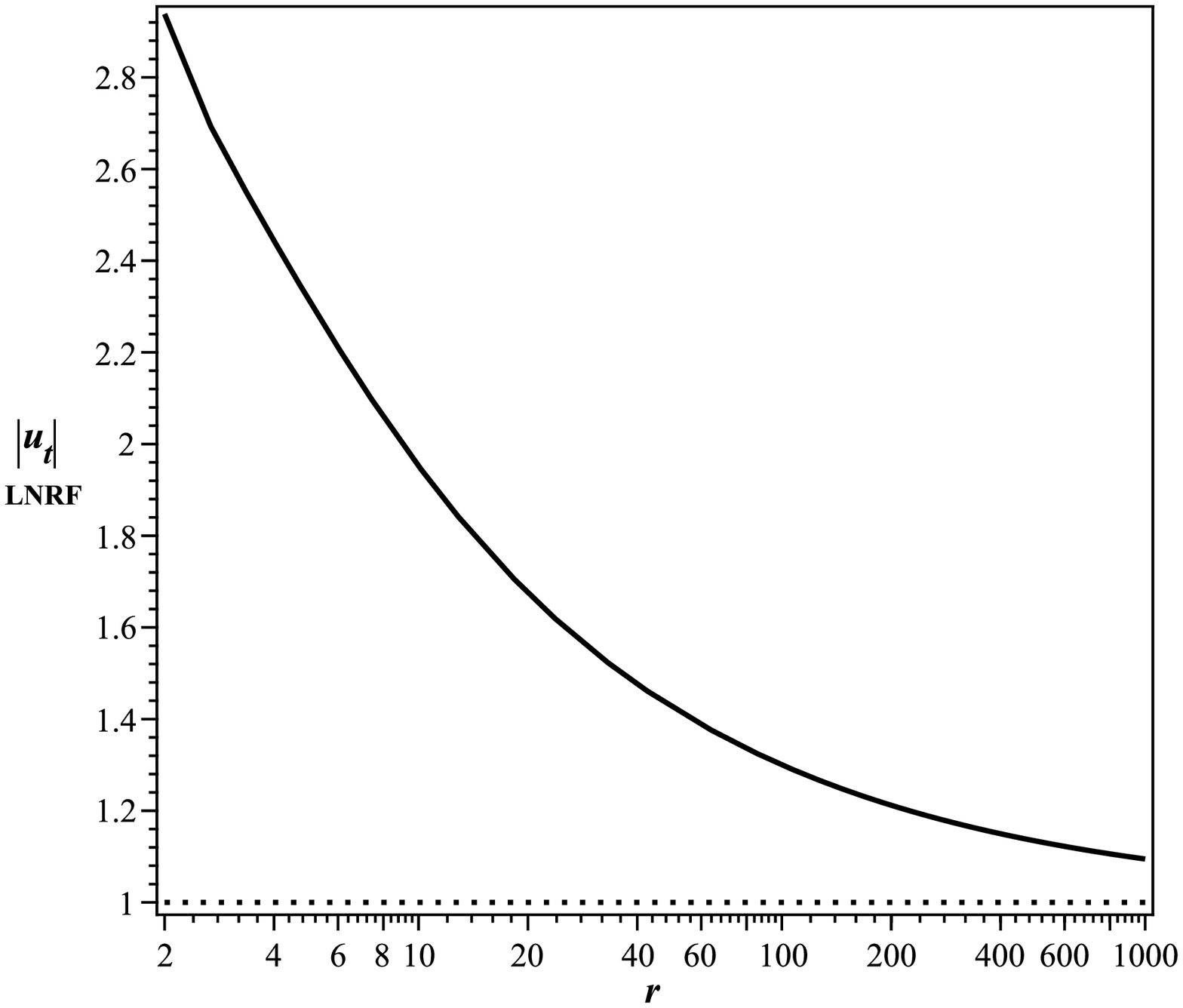} \epsfxsize=3.2in
\epsfysize=2.3in \epsffile{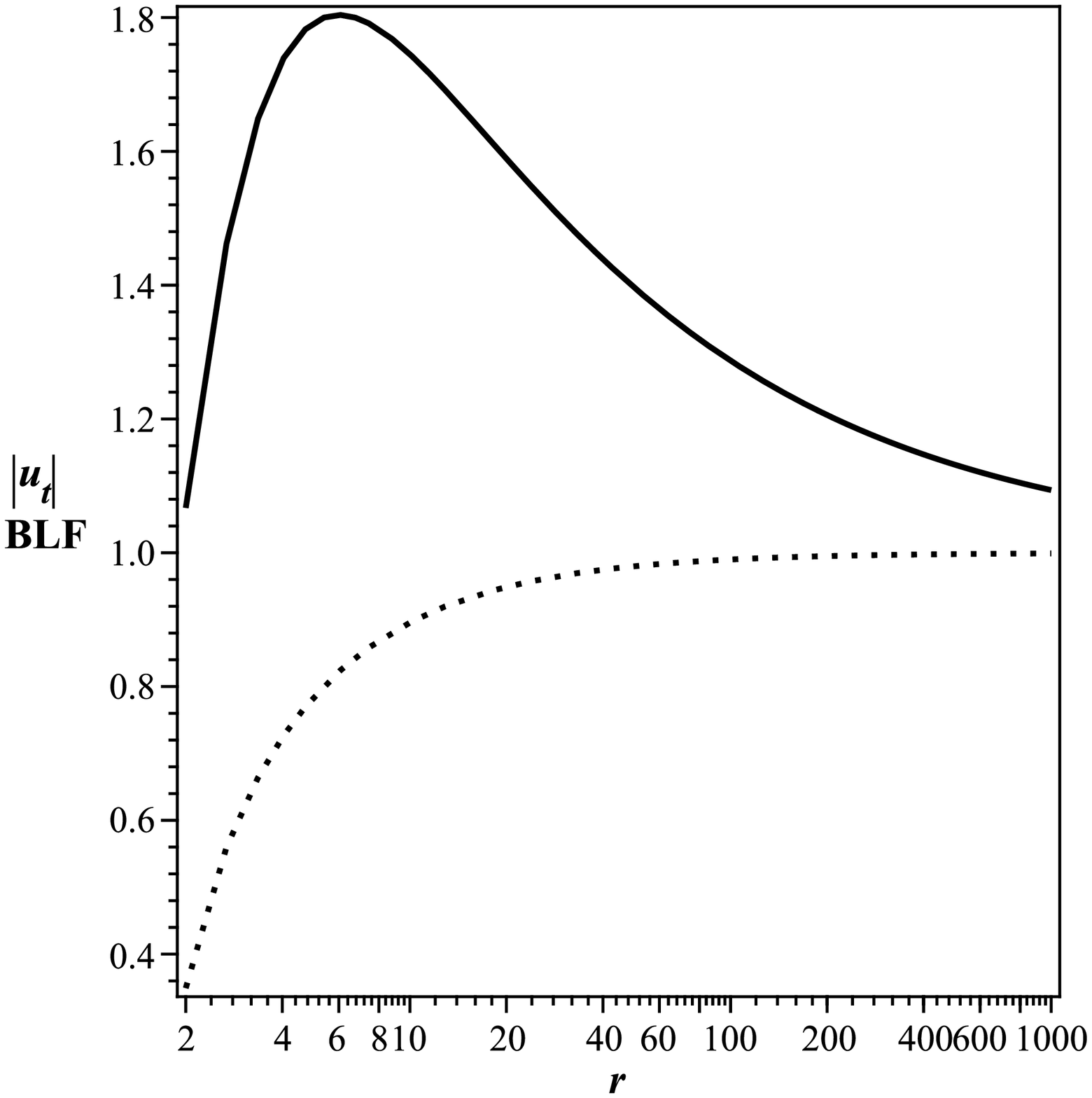}} \epsfxsize=3.2in
\epsfysize=2.3in
\centerline{\epsffile{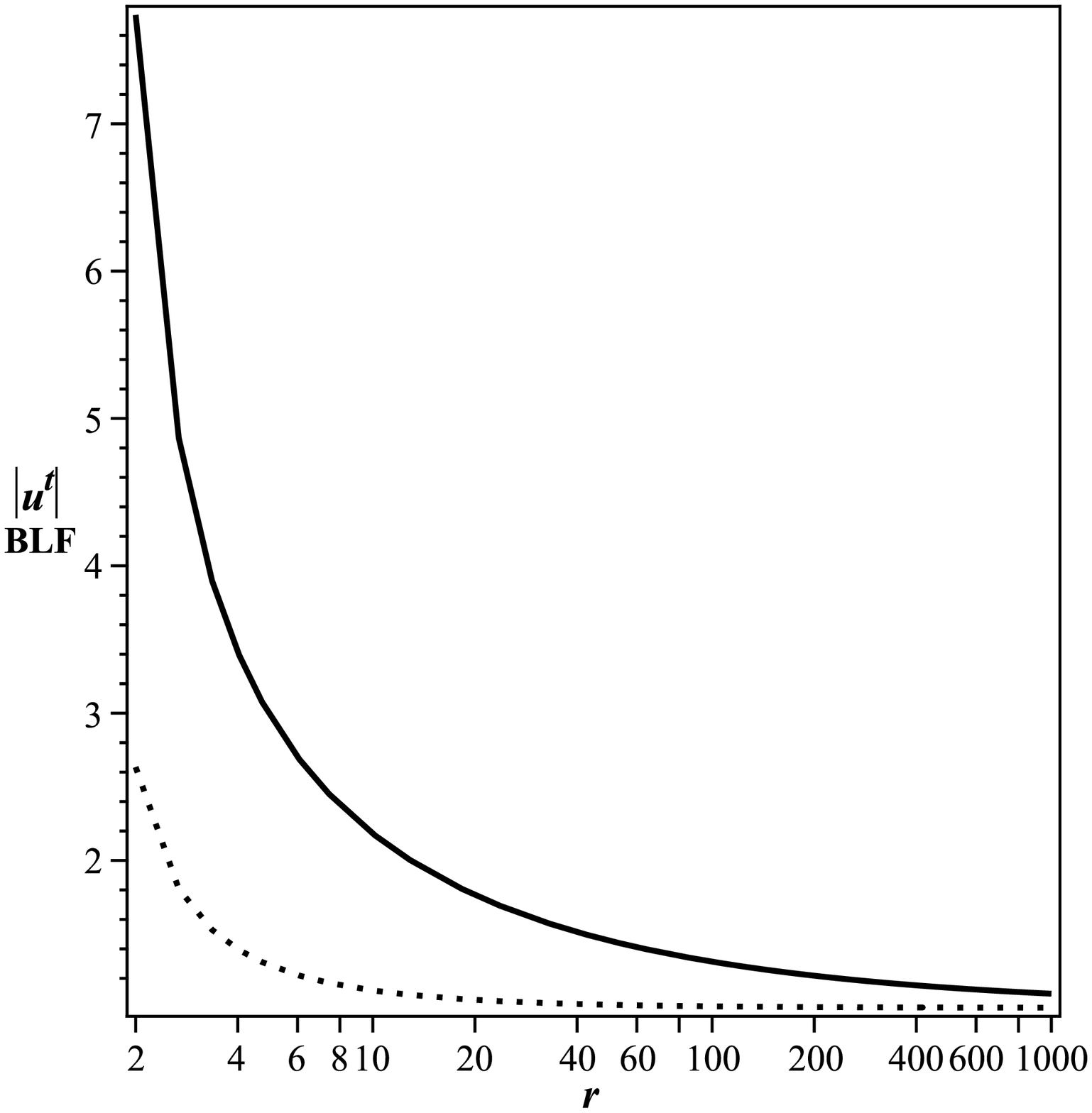}\epsfxsize=3.2in
\epsfysize=2.3in \epsffile{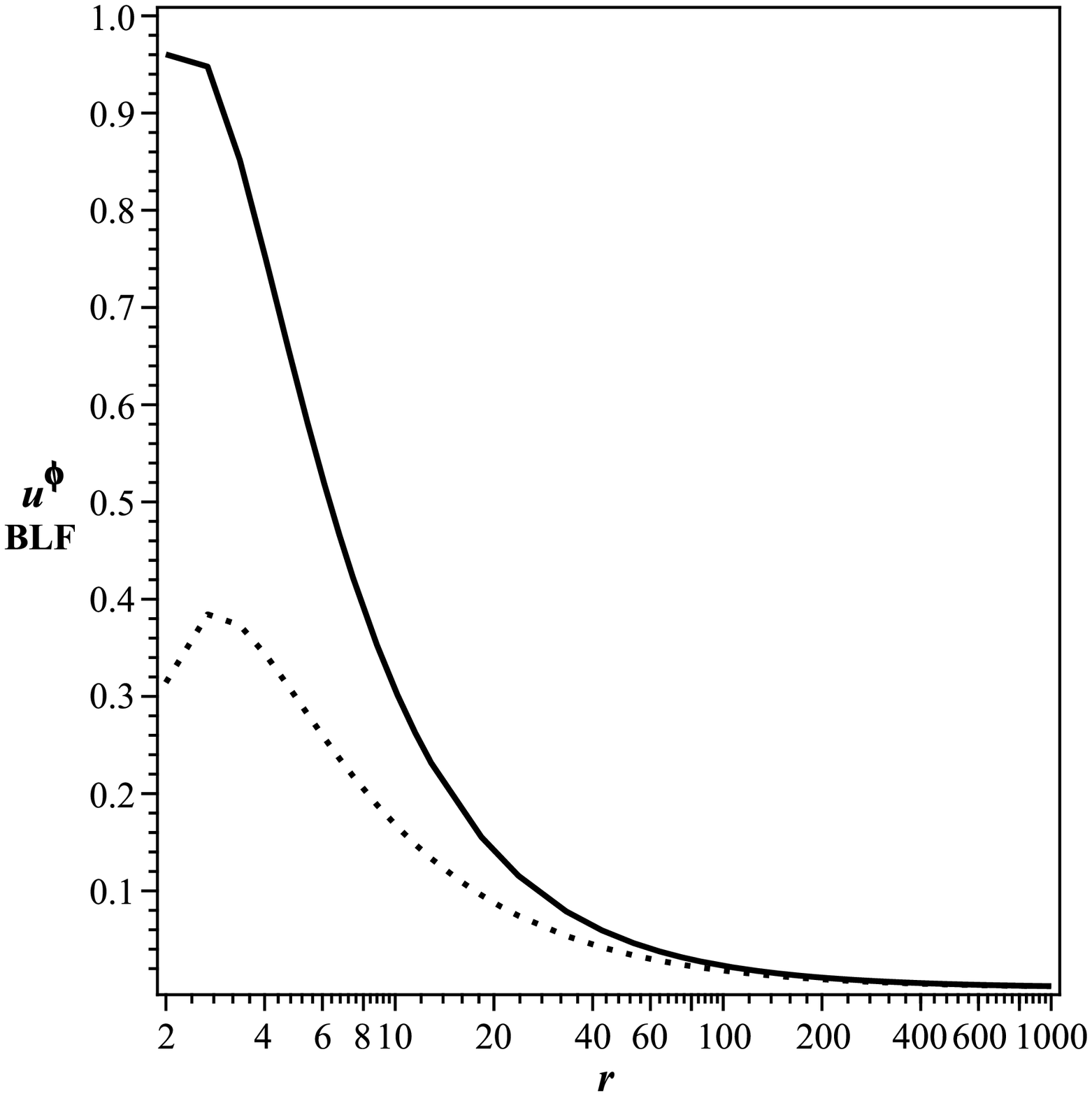}} \epsfxsize=3.2in
\epsfysize=2.3in \centerline{\epsffile{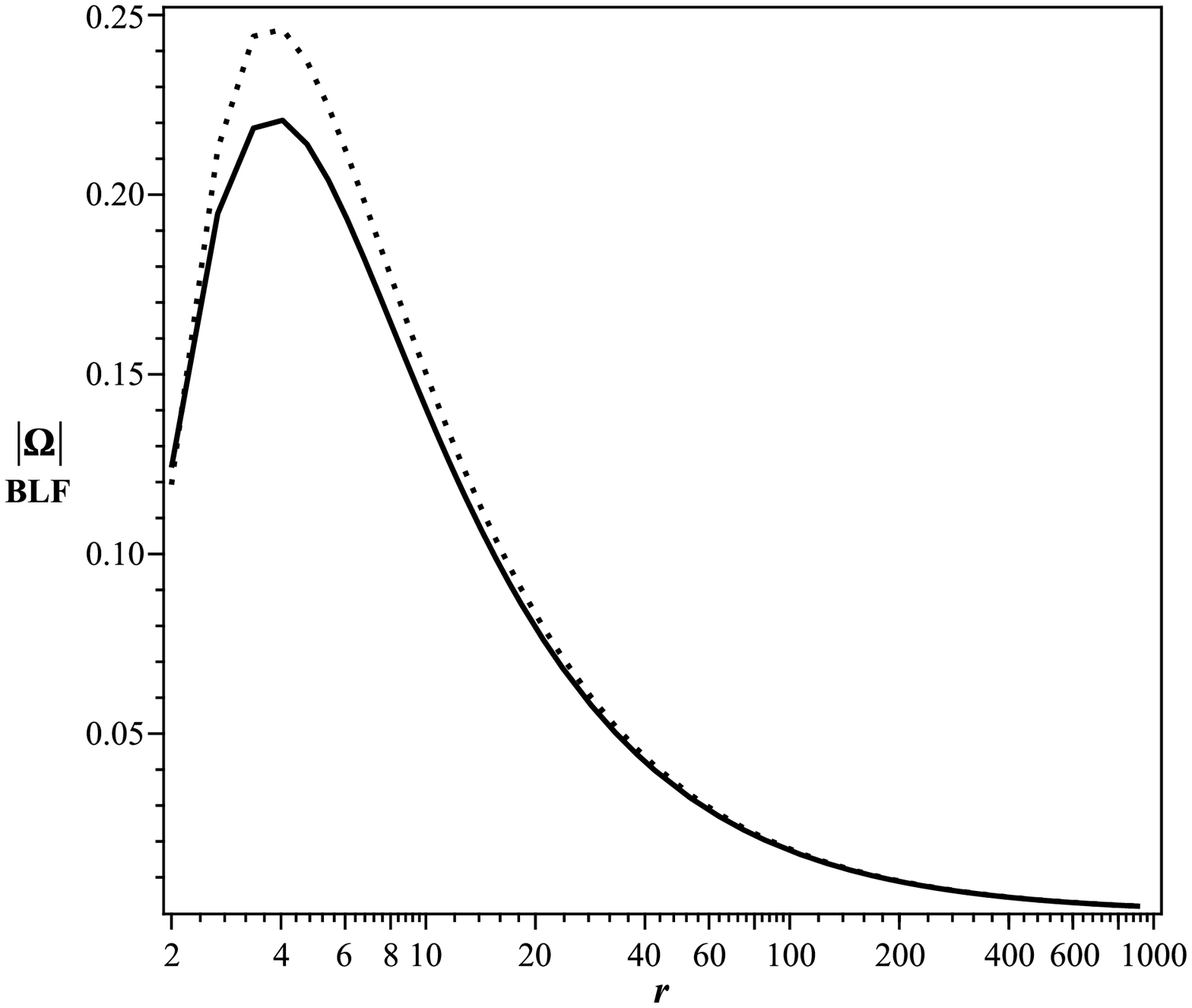}\epsfxsize=3.2in
\epsfysize=2.3in \epsffile{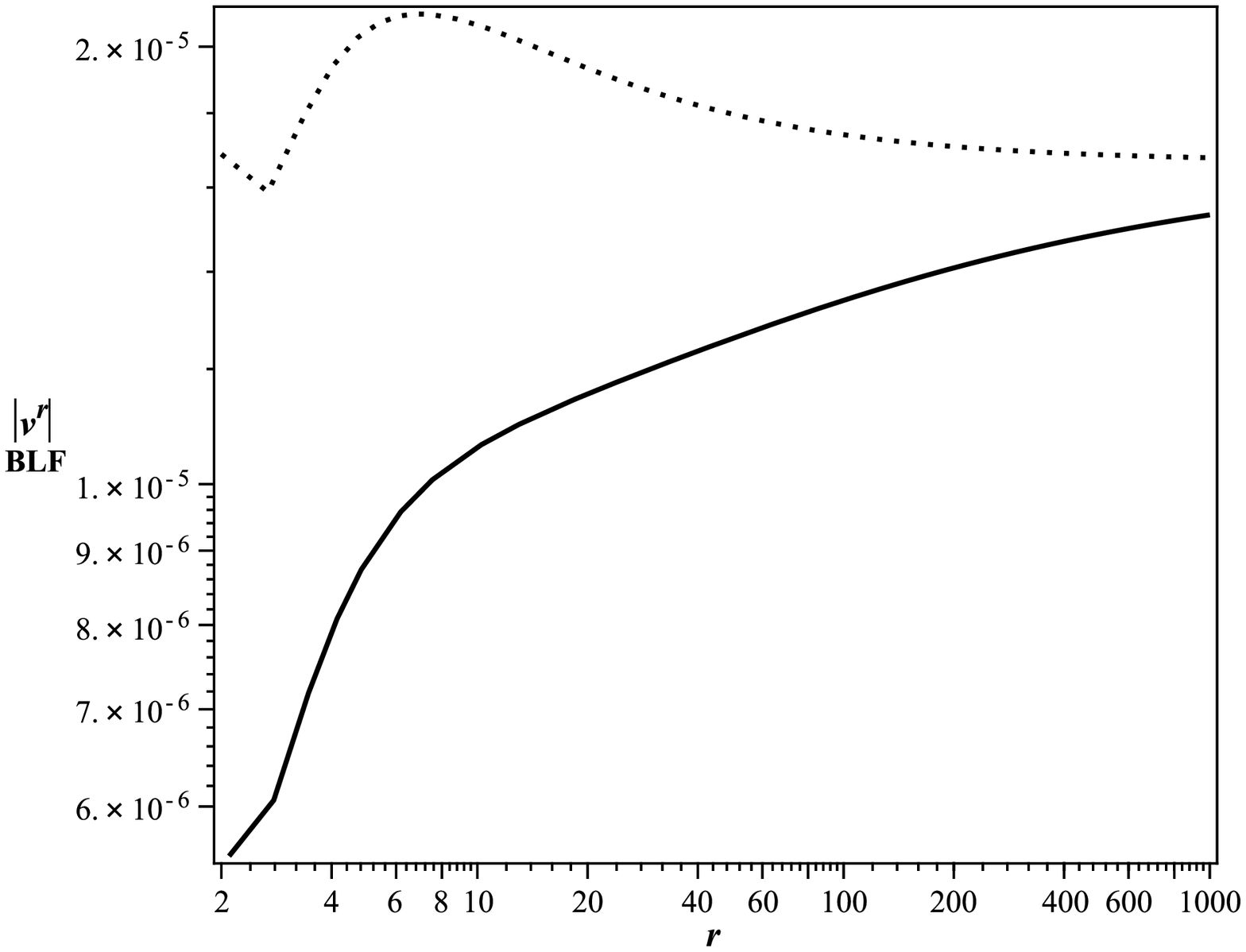}} \caption{\small{Influence of
the $r-t$ component of shear tensor on $\left|u_{\hat{t}}\right|$(in
LNRF) and $\left|u_{t}\right|$, $\left|u^{t}\right|$, $u^{\phi}$(in
BLF), $\left|\Omega\right|$ and $\left|v^{r}\right|$for
$\Omega^{+}$. solid with the $r-t$ component and dotted without
it($\alpha=.01$, $a=.9$, $a_{s}=.1$ and $j=3$).}} \label{figure7}
\end{figure*}
\end{center}
Obviously, $r-t$ component of shear tensor have more influence on
$u_{t}$ and $u^{t}$. Therefore, for
deriving $u_{t}$ and $u^{t}$ we can not vanish the $r-t$ component.
\section{SUMMERY AND CONCLUSION }
In causal viscosity method, only $r-\phi$ component of shear tensor
is assumed to be nonzero in FRF. We do not use this method for
viscosity because our calculations show that there are two nonzero
components of shear tensor in FRF. In our method, we use
azimuthal velocity of LNRF and Keplerian angular velocity, then
we calculate all components of shear tensor in LNRF for two kinds of fluids (rotation in the direction of black
hole ($\Omega^{+}$) and rotation in the opposite direction of black
hole($\Omega^{-}$)). Using transformation tensor we can calculate all
components of shear tensor in BLF and FRF.

Solid curves of figure 1 show unphysical treatments in
shear tensor components close to the inner edge. This unphysical treatment in $\sigma_{\hat{t}\hat{r}}$
may be concerned to assuming $u^{\hat{r}}<<u^{\hat{\phi}}$, therefore if we find a suitable
$v^{\hat{r}}$ this treatment may be resolved. We may have a suitable $v^{\hat{r}}$ if we put $u^{\hat{t}}=\hat{\gamma}=\sqrt{\frac{1}{1-(v^{\hat{\phi}})^2-(v^{\hat{r}})^2}}>0\Rightarrow
1-(v^{\hat{\phi}})^2>(v^{\hat{r}})^2$. According to equation (\ref{23}), the most important origin
of unphysical treatments in $\sigma_{\hat{r}\hat{\phi}}$ is the
$1/\Delta$ term ($1/\Delta$ has a singularity in
the inner edge). Therefor by using Taylor expansion the unphysical
treatment of $\sigma_{\hat{t}\hat{r}}$ and
$\sigma_{\hat{r}\hat{\phi}}$ were solved as it can be seen in dotted curves
of figure 1.

 We solved the hydrodynamical equations in LNRF with assuming
$\eta=1$, we derived the density and four velocity in BLF and LNRF.
The density which is calculated analytically in LNRF is the same in
all frames such as BLF. But comparison of four velocity in LNRF and
BLF (figure 3) shows that $u_{r}$ and $u_{t}$ have a small
difference in two frames. In equation (\ref{37}),
$\sqrt{\Sigma/\Delta}\approx\sqrt{r^2/(r^2-2r+a^2)}$ is near 1,
specially in larger radiuses, therefore $u_{r}$ is similar in two
frames. Following equation (\ref{36}), for $u_{t}$ we have two
parts, the first part is
$(\sqrt{\Delta\Sigma/A}\approx\sqrt{(r-2)/r} )u_{\hat{t}}$ and the
second part is $(-2ar/\sqrt{A\Sigma}\approx-2a/r^2)u_{\hat{\phi}}$.
Because value of $u_{\hat{\phi}}$ is greater than
$u_{\hat{t}}$(dotted curves in figure 3), the first term has more
influence on $u_{t}$ than the second term (the second term is
important just in inner radiuses).

 $u_{\hat{\phi}}$ and $u_{\phi}$ are different in two frames, because in
equation (\ref{38}), $\sqrt{A/\Sigma}\approx r$. The previous equation is logical
because the LNRF rotates with the angular velocity of the frame
dragging (the frame dragging velocity can not be seen in LNRF but it can be seen in BLF), therefore
$u_{\phi}$ is greater than $u_{\hat{\phi}}$ in all radiuses.

In section 5, four velocity was calculated analytically in LNRF and
BLF. The sign of $\Omega$ introduces two types of fluids, $\Omega^{+}$
and $\Omega^{-}$, but it can be seen that for example, in  figures 2 or
4, in $a=.9$ $\Omega^{+}$ is similar to $\Omega^{-}$ in $a=-.9$ (and in
$a=.4$ $\Omega^{+}$ is similar to $\Omega^{-}$ in $a=-.4$ and so on)
therefore we use ($\Omega^{+}$) in figures 5, 6 and 7.

The effects of $\alpha$ coefficient are shown in figure 5. It shows
that if $\alpha$ increases, then the shear viscosity grows, the
radial four velocity will increase and then density decreases.

The $r-t$ component of the shear tensor in LNRF which results from
the relativistic calculations(equation \ref{6})is equal to covariant
derivative of $u_{\hat{t}}$. If we ignore the $r-t$ component of shear
tensor (the dotted curves in figure7), the figures are similar to
the figures of the previous works such as Gammie \& Popham (1998)
and Takahashi (2007b). But, with $r-t$ component of shear tensor
(the solid curves in figure 7) a few number of four velocities
change in all frames, specially $u^{\hat{t}}$, $u^{t}$, $u^{\phi}$
and $v^{r}$ will change more than the others. The influences of the
$r-t$ component on all physical variables can be derived if we solve
all the equations of the disk with a state equation numerically.

When we have time dilation $u^{t}$ has to be greater than
1($dt>d\tau\Rightarrow u^{t}=\frac{dt}{d\tau}>1$). According to
figure 4, the effect of time dilation is much greater in  the inner
edge than the other places. If we ignore the $r-t$ component of
shear tensor or put $\left|u_{\hat{t}}\right|=1$(dotted curve in the
first panel of first column of figure 7), it is equivalent to
ignoring time dilation.  Equations (\ref{35'}) and (\ref{37}) show
that the $r-t$ component of shear tensor have no influence on
$u^{\hat{r}}$ and $u^{r}$. Near the black hole $u^{t}$ increases,
which causes a decreases in $v^{r}$ ($v^{r}=\frac{u^{r}}{u^{t}}$).

Energy can be calculated from $T^{tt}=\rho\eta u^{t}u^{t}+pg^{tt}$,
where the first term of energy is related to $u_{t}$. Figure 2 shows
that $\rho$ is greater in inner edge than the other places and also
is the same with the $r-t$ component of shear tensor and without it.
The $u_{t}$ which uses the $r-t$ component of shear tensor is
greater than the $u_{t}$ without it. Therefore, the first term of
energy with $r-t$ component of shear tensor is greater than the this
term of energy without $r-t$ component.

 Far from the black hole the general relativistic influences are vanished,
therefore in outer edge, the influences of $r-t$ component of shear
tensor are too low (as it is seen in outer radiuses of figure 7).

If we want to calculate the temperature, pressure, internal energy,
cooling and heating or radiation, we must use a suitable state
equation with equations (\ref{11}), (\ref{14}), (\ref{16}) and
(\ref{20})(energy equation), then, to solve the equations
numerically suitable boundary conditions are needed.\\\\
\textbf{acknowledgments}
We are grateful to the referee
for a very careful reading of the manuscript and for
his/her suggestions, which have helped us improve the
presentation of our results.


\appendix

\section{Metric Components}
Nonzero components of Kerr metric are given as:
\begin{eqnarray}
&g_{tt}&=-\alpha^{2}+\beta_{\phi}\beta^{\phi}=-(1-\frac{2mr}{\Sigma}),\qquad    g_{rr}=\gamma_{rr}=\frac{\Sigma}{\Delta},\qquad \nonumber\\
&
g_{t\phi}&=\beta_{\phi}=-\frac{2mar\sin^2
\theta}{\Sigma},\qquad g_{\theta\theta}=\gamma_{\theta\theta}=\Sigma,\nonumber\\
&g_{\phi\phi}&=\gamma_{\phi\phi}=\frac{A\sin^2
\theta}{\Sigma},
\end{eqnarray}
and inverse components of metric $g^{\mu\nu}$ are:
\begin{eqnarray}
&g^{tt}&=-\frac{1}{\alpha^2}=-\frac{A}{\Sigma\Delta},\qquad g^{t\phi}=\frac{\beta^{\phi}}{\alpha^2}=-\frac{2mar}{\Sigma\Delta},\nonumber\\ &g^{rr}&=\gamma^{rr}=\frac{\Delta}{\Sigma},\qquad g^{\theta\theta}=\gamma^{\theta\theta}=\frac{1}{\Sigma} ,\nonumber\\  &g^{\phi\phi}&=\gamma^{\phi\phi}-\frac{(\beta^{\phi})^2}{\alpha^2}=\frac{1
}{\Delta\sin^2\theta}(1-\frac{2mr}{\Sigma}).
\end{eqnarray}

\section{Transformation Between BLF, LNRF and FRF}
Components of $e_{\mu}^{\hat{\nu}}$ connecting between BLF(
Boyer-Lindquist Frame) and LNRF(Locally non-rotating Frame) are
calculated as:
\begin{eqnarray}
&&\left(%
\begin{array}{cccc}
   e_{t}^{\hat{t}} & e_{t}^{\hat{r}} & e_{t}^{\hat{\theta}} & e_{t}^{\hat{\phi}}\\
 e_{r}^{\hat{t}} & e_{r}^{\hat{r}} & e_{r}^{\hat{\theta}} & e_{r}^{\hat{\phi}} \\
  e_{\theta}^{\hat{t}} & e_{\theta}^{\hat{r}} & e_{\theta}^{\hat{\theta}} & e_{\theta}^{\hat{\phi}} \\
  e_{\phi}^{\hat{t}} & e_{\phi}^{\hat{r}} & e_{\phi}^{\hat{\theta}} & e_{\phi}^{\hat{\phi}} \\
\end{array}%
\right)=\nonumber\\ &&\left(%
\begin{array}{cccc}
  (\frac{\Sigma\Delta}{A})^\frac{1}{2} & 0 & 0 & -\frac{2Mar\sin\theta}{(\Sigma A)^\frac{1}{2}} \\
  0 & (\frac{\Sigma}{\Delta})^\frac{1}{2} & 0 & 0 \\
  0 & 0 & \Sigma^\frac{1}{2} & 0 \\
  0 & 0 & 0 & (\frac{A}{\Sigma})^\frac{1}{2}\sin \theta \\
\end{array}%
\right),
\end{eqnarray}
\begin{eqnarray}
&&\left(%
\begin{array}{cccc}
   e^{t}_{\hat{t}} & e^{t}_{\hat{r}} & e^{t}_{\hat{\theta}} & e^{t}_{\hat{\phi}}\\
 e^{r}_{\hat{t}} & e^{r}_{\hat{r}} & e^{r}_{\hat{\theta}} & e^{r}_{\hat{\phi}} \\
  e^{\theta}_{\hat{t}} & e^{\theta}_{\hat{r}} & e^{\theta}_{\hat{\theta}} & e^{\theta}_{\hat{\phi}} \\
  e^{\phi}_{\hat{t}} & e^{\phi}_{\hat{r}} & e^{\phi}_{\hat{\theta}} & e^{\phi}_{\hat{\phi}} \\
\end{array}%
\right)=\nonumber\\&&\left(%
\begin{array}{cccc}
  (\frac{A}{\Sigma\Delta})^\frac{1}{2} & 0 & 0 & \frac{2Mar}{(\Sigma A\Delta)^\frac{1}{2}} \\
  0 & (\frac{\Delta}{\Sigma})^\frac{1}{2} & 0 & 0 \\
  0 & 0 & \frac{1}{\Sigma^\frac{1}{2}} & 0 \\
  0 & 0 & 0 & (\frac{\Sigma}{A})^\frac{1}{2}\frac{1}{\sin \theta} \\
\end{array}%
\right).
\end{eqnarray}
The transformation between LNRF and FRF(Fluid Rest Frame) are as
follows:
\begin{eqnarray}
&&\left(%
\begin{array}{cccc}
e^{\hat{t}}_{(t)} & e^{\hat{t}}_{(r)} & e^{\hat{t}}_{(\theta)} & e^{\hat{t}}_{(\phi)}\\
 e^{\hat{r}}_{(t)} & e^{\hat{r}}_{(r)} & e^{\hat{r}}_{(\theta)} & e^{\hat{r}}_{(\phi)} \\
  e^{\hat{\theta}}_{(t)} & e^{\hat{\theta}}_{(r)} & e^{\hat{\theta}}_{(\theta)} & e^{\hat{\theta}}_{(\phi)} \\
  e^{\hat{\phi}}_{(t)} & e^{\hat{\phi}}_{(r)} & e^{\hat{\phi}}_{(\theta)} & e^{\hat{\phi}}_{(\phi)} \\\end{array}%
\right)=\nonumber\\&&\left(%
\begin{array}{cccc}
  \hat{\gamma} & \hat{\gamma}v_{\hat{r}} & 0 & \hat{\gamma}v_{\hat{\phi}} \\
  \hat{\gamma}v_{\hat{r}} &1+\frac{\hat{\gamma}^2v_{\hat{r}}^{2}}{1+\hat{\gamma}} & 0 & \frac{\hat{\gamma}^2v_{\hat{r}}v_{\hat{\phi}}}{1+\hat{\gamma}} \\
  0 & 0 & 1 & 0 \\
  \hat{\gamma}v_{\hat{\phi}} & \frac{\hat{\gamma}^2v_{\hat{r}}v_{\hat{\phi}}}{1+\hat{\gamma}} & 0 & 1+\frac{\hat{\gamma}^2v_{\hat{\phi}}^{2}}{1+\hat{\gamma}} \\
\end{array}%
\right),
\end{eqnarray}
\begin{eqnarray}
&&\left(%
\begin{array}{cccc}
  e_{\hat{t}}^{(t)} & e_{\hat{t}}^{(r)} & e_{\hat{t}}^{(\theta)} & e_{\hat{t}}^{(\phi)}\\
 e_{\hat{r}}^{(t)} & e_{\hat{r}}^{(r)} & e_{\hat{r}}^{(\theta)} & e_{\hat{r}}^{(\phi)} \\
  e_{\hat{\theta}}^{(t)} & e_{\hat{\theta}}^{(r)} & e_{\hat{\theta}}^{(\theta)} & e_{\hat{\theta}}^{(\phi)} \\
  e_{\hat{\phi}}^{(t)} & e_{\hat{\phi}}^{(r)} & e_{\hat{\phi}}^{(\theta)} & e_{\hat{\phi}}^{(\phi)} \\
\end{array}%
\right)=\nonumber\\&&\left(%
\begin{array}{cccc}
  e^{\hat{t}}_{(t)} & -e^{\hat{t}}_{(r)} & -e^{\hat{t}}_{(\theta)} & -e^{\hat{t}}_{(\phi)} \\
  - e^{\hat{r}}_{(t)} & e^{\hat{r}}_{(r)} & e^{\hat{r}}_{(\theta)} & e^{\hat{r}}_{(\phi)} \\
  -e^{\theta}_{(t)} & e^{\hat{\theta}}_{(r)} & e^{\hat{\theta}}_{(\theta)} & e^{\hat{\theta}}_{(\phi)} \\
  -e^{\hat{\phi}}_{(t)} & e^{\hat{\phi}}_{(r)} & e^{\hat{\phi}}_{(\theta)} & e^{\hat{\phi}}_{(\phi)} \\
\end{array}%
\right).
\end{eqnarray}
Where we use $u^{\hat{t}}=-u_{\hat{t}}=\alpha u^{t}$, then the Lorentz factor $\hat{\gamma}$ is calculated as:
\begin{equation}
\hat{\gamma}\equiv (1-\hat{v}^2)^{-\frac{1}{2}}=\alpha
u^{t},\qquad (\hat{v^2}=v_{\hat{i}}v^{\hat{i}}=v_{\hat{r}}^2+v_{\hat{\theta}}^2+v_{\hat{\phi}}^2).
\end{equation}
In LNRF, three velocity components are calculated as:
\begin{equation}
v^{\hat{i}}=\frac{u^{\hat{i}}}{u^{\hat{t}}},\qquad (i=r,\theta,\phi).
\end{equation}
After some calculation we have ($\Omega=\frac{u^{\phi}}{u^{t}}$):
\begin{equation}
v^{\hat{r}}=\frac{A^\frac{1}{2}}{\Delta}\frac{u^{r}}{u^{t}},v^{\hat{\theta}}=0,\qquad v^{\hat{\phi}}=\frac{\sqrt{\gamma_{\phi\phi}}}{\alpha}(\beta^{\phi}+\Omega).
\end{equation}

\section{Deriving Four Velocity in LNRF}
From equation (\ref{27}) we have:
\begin{equation}
v^{\hat{\phi}}\pm =\frac{A}{r^{2}\sqrt{\Delta}}(\frac{\pm1}{r^{3/2}\pm a}-\frac{2ar}{A}).
\end{equation}
First we derive four velocity for $\Omega_{k}+(\Omega^{+})=1/(r^\frac{3}{2}+a)$ as
\begin{equation}
(v^{\hat{\phi}})^2=\frac{A^2+4a^2r^2(r^\frac{3}{2}+a)^2-4arA(r^\frac{3}{2}+a)}{r^4(r^\frac{3}{2}+a)^2\Delta}.
\end{equation}
In LNRF $\hat{\gamma}=u^{\hat{t}}=\sqrt{\frac{1}{1-\hat{v}^{2}}}$
and in our study we put $u_{\theta}=0$. If we suppose
$u_{\hat{\phi}}>>u_{\hat{r}}$ (as it was in pervious papers such as Popham \&
Gammie 1998 and Takahashi 2007a,b), then $\hat{v^{2}}\approx
(v^{\hat{\phi}})^2$ and we have:
\begin{eqnarray}
&&\hat{\gamma}=u^{\hat{t}}=-u_{\hat{t}}=r\sqrt{\Delta}(r^\frac{3}{2}+a)\times\nonumber\\&&(r^4(r^\frac{3}{2}+a)^2\Delta-A^2-4a^2r^2(r^\frac{3}{2}+a)^2+4arA(r^\frac{3}{2}+a))^{-\frac{1}{2}},\nonumber\\
\end{eqnarray}
also we know that $v^{\hat{\phi}}=u^{\hat{\phi}}/u^{\hat{t}}$
therefore,
\begin{eqnarray}
&&u^{\hat{\phi}}=u^{\hat{t}}v^{\hat{\phi}}=(r^{3}+ra^{2}-2ar^{\frac{3}{2}})\times\nonumber\\&&(r^4(r^\frac{3}{2}+a)^2\Delta-A^2-4a^2r^2(r^\frac{3}{2}+a)^2+4arA(r^\frac{3}{2}+a))^{-\frac{1}{2}}.\nonumber\\
\end{eqnarray}
For $\Omega_{k}^{-}(\Omega^{-})=-1/(r^\frac{3}{2}-a)$ we have:
\begin{eqnarray}
&&\hat{\gamma}=u^{\hat{t}}=-u_{\hat{t}}=\frac{1}{\sqrt{1-(v^{\hat{\phi}})^2}}= r\sqrt{\Delta}(r^\frac{3}{2}-a)\times\nonumber\\&&(r^4(r^\frac{3}{2}-a)^2\Delta-A^2-4a^2r^2(r^\frac{3}{2}-a)^2-4arA(r^\frac{3}{2}-a))^{-\frac{1}{2}},
\nonumber\\&&
u^{\hat{\phi}}=u^{\hat{t}}  v^{\hat{\phi}}=\-(r^{3}+ra^{2}+2ar^{\frac{3}{2}})\times\nonumber\\&&(r^4(r^\frac{3}{2}-a)^2\Delta-A^2-4a^2r^2(r^\frac{3}{2}-a)^2-4arA(r^\frac{3}{2}-a))^{-\frac{1}{2}}.\nonumber\\
\end{eqnarray}
\section{Nonzero components of shear tensor in BLF and FRF}
Nonzero components of shear tensor in BLF can be calculated by $\sigma_{\alpha\beta}=e^{\hat{\mu}}_{\alpha}e^{\hat{\nu}}_{\beta}\sigma_{\hat{\mu}\hat{\nu}}$ where $e^{\hat{\mu}}$ and $e^{\hat{\nu}}$ are given in Appendix B. Therefore, these components for $\Omega^{+}$ are:
\begin{eqnarray}\label{a}
&&\sigma^{+}_{tr}=\sigma^{+}_{rt}=\frac{1}{4B^{+}\Delta^{\frac{1}{2}}}(6r^\frac{9}{2}a^6+8r^8a^3+36r^\frac{15}{2}a^2+4r^\frac{7}{2}a^6\nonumber\\&&-24r^\frac{11}{2}a^4-4r^4a^5-40r^7a^3-8r^5a^5+2r^{10}a+6r^6a^5\nonumber\\&&-18r^\frac{17}{2}a^2-4r^\frac{13}{2}a^4+36r^6a^3)\nonumber\\&&
+\frac{1}{4B^{+}\Delta^{\frac{3}{2}}}(+3a^8r^\frac{11}{2}+r^\frac{27}{2}+42r^{11}a-12a^7r^6+2r^\frac{9}{2}a^8\nonumber\\&&+6r^\frac{23}{2}a^2+12r^\frac{19}{2}a^4-12r^{12}a-36a^5r^8+54r^7a^5-6r^5a^7\nonumber\\&&+12a^5r^6-36r^{10}a-40a^4r^\frac{15}{2}-72r^8a^3-36r^{10}a^3-2r^\frac{25}{2}\nonumber\\&&-56r^\frac{19}{2}a^2-8r^\frac{13}{2}a^4+10r^\frac{15}{2}a^6+72r^\frac{17}{2}a^2+102r^9a^3),\nonumber\\
&&\sigma^{+}_{r\phi}=\sigma^{+}_{\phi r}=-\frac{\sqrt{A}}{4B^{+}\sqrt{\Delta}}(3a^5r^\frac{7}{2}+2a^5r^\frac{5}{2}+3a^4r^5-9r^\frac{15}{2}a\nonumber\\&&+18r^5a^2-12r^\frac{9}{2}a^3+18r^\frac{13}{2}a-2r^\frac{11}{2}a^3-2r^3a^4-20r^6a^2\nonumber\\&&-4r^4a^4+4r^7a^2+r^9).
\end{eqnarray}
Non-zero components of shear tensor in FRF can be calculated by
$\sigma_{(\alpha)(\beta)}=e^{\hat{\mu}}_{(\alpha)}e^{\hat{\nu}}_{(\beta)}\sigma_{\hat{\mu}\hat{\nu}}$
where $e^{\hat{\mu}}_{(\alpha)}$ and $e^{\hat{\nu}}_{(\beta)}$ are
transformation matrixes of Appendix B.
\begin{eqnarray}\label{cc}
&&\sigma_{(\alpha)(\beta)}=
\left(%
\begin{array}{cccc}
  \hat{\gamma} & 0 & 0 & u^{\hat{\phi}} \\
  0 & 1 & 0 & 0 \\
  0 & 0 & 1 & 0 \\
  u^{\hat{\phi}} & 0 & 0 & 1+\frac{u_{\hat{\phi}}^{2}}{1+\hat{\gamma}} \\
\end{array}%
\right)\times\nonumber\\&&
\left(%
\begin{array}{cccc}
  \hat{\gamma} & 0 & 0 & u_{\hat{\phi}} \\
  0 & 1 & 0 & 0 \\
  0 & 0 & 1 & 0 \\
  u_{\hat{\phi}} & 0 & 0 & 1+\frac{u_{\hat{\phi}}^{2}}{1+\hat{\gamma}} \\
\end{array}%
\right)\times
\left(%
\begin{array}{cccc}
  0 & \sigma_{\hat{r}\hat{t}} & 0 & 0 \\
  \sigma_{\hat{r}\hat{t}} & 0 & 0 & \sigma_{\hat{r}\hat{\phi}} \\
  0 & 0 & 0 & 0 \\
  0 & \sigma_{\hat{r}\hat{\phi}} & 0 & 0 \\
\end{array}%
\right).\nonumber\\
\end{eqnarray}
Also, we have:

\begin{eqnarray}\label{c}
&&\left(
\begin{array}{cccc}
  \sigma_{(t)(t)} & \sigma_{(t)(r)} & \sigma_{(t)(\theta)} & \sigma_{(t)(\phi)} \\
  \sigma_{(r)(t)} & \sigma_{(r)(r)} & \sigma_{(r)(\theta)} & \sigma_{(r)(\phi)} \\
  \sigma_{(\theta)(t)} & \sigma_{(\theta)(r)} & \sigma_{(\theta)(\theta)} & \sigma_{(\theta)(\phi)} \\
  \sigma_{(\phi)(t)} & \sigma_{(\phi)(r)} & \sigma_{(\phi)(\theta)} & \sigma_{(\phi)(\phi)} \\
\end{array}%
\right)=\nonumber\\
&&\left(
\begin{array}{cccc}
  0 & u_{\hat{\phi}}\sigma_{\hat{r}\hat{\phi}} & 0 & 0 \\
  u_{\hat{\phi}}\sigma_{\hat{r}\hat{\phi}} & 0 & 0 & \sigma_{\hat{r}\hat{t}}u_{\hat{\phi}}+\sigma_{\hat{r}\hat{\phi}}+\frac{\sigma_{\hat{r}\hat{\phi}}u_{\hat{\phi}}^{2}}{1+\hat{\gamma}} \\
  0 & 0 & 0 & 0 \\
  0 & \sigma_{\hat{r}\hat{t}}u_{\hat{\phi}}+\sigma_{\hat{r}\hat{\phi}}+\frac{\sigma_{\hat{r}\hat{\phi}}u_{\hat{\phi}}^{2}}{1+\hat{\gamma}} & 0 & 0 \\
\end{array}%
\right),\nonumber\\&&\nonumber\\&&
\sigma_{(t)(r)}=\sigma_{(r)(t)}=u_{\hat{\phi}}\sigma_{\hat{r}\hat{\phi}},\nonumber\\&&
\sigma_{(r)(\phi)}=\sigma_{(\phi)(r)}=\sigma_{\hat{r}\hat{t}}u_{\hat{\phi}}+\sigma_{\hat{r}\hat{\phi}}+\frac{\sigma_{\hat{r}\hat{\phi}}u_{\hat{\phi}}^{2}}{1+\hat{\gamma}}.
\end{eqnarray}
Therefore, in FRF, two components of shear tensor are non-zero.


\label{lastpage}


\begin{thebibliography}{}
  \bibitem[\protect\citeauthoryear{Abramowicz et al.}{1992}]{bu}
    Abramowicz, M. A., Chen, X., Granath, M., Lasota, J.-P., 1996, ApJ, 471, 762
  \bibitem[\protect\citeauthoryear{The Chicago Manual}%
    {1982}]{ch} Abramowicz, M. A., Lanza, A., Percival, M. J., 1997, ApJ, 479, 179
  \bibitem[\protect\citeauthoryear{ Bardeen}{1970}]{bl}
    Bardeen, J. M., 1970, ApJ, 162, 71
  \bibitem[\protect\citeauthoryear{Bardeen et al.}{1972}]{ed}
    Bardeen, J. M., Press, W. H., Teukolsky, S. A., 1972, ApJ, 178, 347
  \bibitem[\protect\citeauthoryear{Chakrabarti}{1986}]{la}
    Chakrabarti, S., 1996, ApJ, 471, 237
  \bibitem[\protect\citeauthoryear{Frolov \& Novikov}%
    {1989}]{ms} Frolov, V. P., Novikov, I. D., 1998, Black Hole Physics: Basic Concepts and New Developments, Kluwer Academic
  \bibitem[\protect\citeauthoryear{Gammie \& Popham}{1998}]
     {GP} Gammie, C. F.,Popham, R., 1998, ApJ, 498, 313
  \bibitem[\protect\citeauthoryear{Lasota}{1994}]
    {la} Lasota, J.-P. 1994, in Theory of Accretion Disks-2, ed. W. J. Duschl, J. Frank, F. Meyer, E. Meyer-Hofmeister, W. M. Tscharnuter
  \bibitem[\protect\citeauthoryear{Manmoto}{2000}]
    {ma} Manmoto, T., 2000, ApJ, 534, 734
  \bibitem[\protect\citeauthoryear{Misner et al.}%
    {1973}]{mtw} Misner C. W., Thorne K. S.,
      Wheeler J. A., 1973, Gravitation.
      Freeman, San Francisco
  \bibitem[\protect\citeauthoryear{Papaloizou\& Szuszkiewicz}
      {1994}]{sa} Papaloizou, J. C. B.,Szuszkiewicz, E., 1994, MNRAS, 268, 29
  \bibitem[\protect\citeauthoryear{ Peitz, J.,Appl, S., 1997, MNRAS, 286, 681}%
      {1997}]{pa} Peitz, J., Appl, S., 1997a, MNRAS, 286, 681
  \bibitem[\protect\citeauthoryear{ Peitz \& Appl}%
     {1997}]{paa} Peitz, J., Appl, S., 1997b, MNRAS, 296, 231
   \bibitem[\protect\citeauthoryear{Popham \& Gammie}%
     {1998}]{pg}Popham, R., Gammie, C. F., 1998, ApJ, 504, 419
  \bibitem[\protect\citeauthoryear{Riffert \& Herold,}%
     {1998}]{rh}Riffert, H., Herold, H. 1995, ApJ, 450, 508
  \bibitem[\protect\citeauthoryear{Shakura \& Sunyaev}
     {1973}]{ss}Shakura, N. I., Sunyaev, R. A. 1973, A\& A, 24, 337
  \bibitem[\protect\citeauthoryear{Takahashi}%
     {2007}]{ta}Takahashi, R., 2007a, A\& A, 461, 393
  \bibitem[\protect\citeauthoryear{Takahashi}%
     {2007}]{tb}Takahashi, R., 2007b, MNRAS, 382, 567 %
    \end{thebibliography}
\end{document}